\documentclass[journal, 10pt]{IEEEtran}
\IEEEoverridecommandlockouts

\usepackage{color}
\usepackage{theorem}
\usepackage{times,amsmath,epsfig}
\usepackage{amssymb}
\usepackage{subfigure}
\usepackage{cite}
\usepackage{hyperref}
\usepackage{algorithmic}
\usepackage{algorithm}
\usepackage{needspace}

\usepackage{tikz}
\usetikzlibrary{shapes,arrows}
\input{mysymbol.sty}
\tikzstyle{phantom vertex} = [ ellipse, 
                               anchor = center, 
                               minimum height = 1*\unit, 
                               minimum width  = 1*\unit,
                               inner sep=0pt,
                               anchor=center]
\tikzstyle{red vertex}   = [black, fill = red!20,   phantom vertex, draw]
\tikzstyle{black vertex} = [black, fill = black!20, phantom vertex, draw]
\tikzstyle{blue vertex}  = [black, fill = blue!20,  phantom vertex, draw]
\tikzstyle{green vertex} = [black, fill = green!20,  phantom vertex, draw]
\tikzstyle{yellow vertex} = [black, fill = yellow!20,  phantom vertex, draw]
\tikzstyle{cyan vertex} = [black, fill = cyan!20,  phantom vertex, draw]
\tikzstyle{vertex}       = [draw, phantom vertex]

\tikzstyle{point} = [ellipse, inner sep=0pt, draw, fill=white, anchor = center,
                     minimum height = 0.05*\unit, minimum width  = 0.05*\unit]

\newcommand{\QED}{\hfill\ensuremath{\blacksquare}}
\newcommand{\vvec}{{\mathrm{vec}}}

\newtheorem{problem}{\bf Problem}

\def \bbCo {\bbC_{{\scriptscriptstyle \ccalO}}}
\def \bbCh {\bbC_{{\scriptscriptstyle \ccalH}}}
\def \bbCoh {\bbC_{{\scriptscriptstyle \ccalO\ccalH}}}

\def \hbCo {\hbC_{{\scriptscriptstyle \ccalO}}}
\def \hbCh {\hbC_{{\scriptscriptstyle \ccalH}}}
\def \hbCoh {\hbC_{{\scriptscriptstyle \ccalO\ccalH}}}
\def \hbCho {\hbC_{{\scriptscriptstyle \ccalH\ccalO}}}

\def \bbSo {\bbS_{{\scriptscriptstyle \ccalO}}}
\def \bbSh {\bbS_{{\scriptscriptstyle \ccalH}}}
\def \bbSoh {\bbS_{{\scriptscriptstyle \ccalO\ccalH}}}
\def \bbSho {\bbS_{{\scriptscriptstyle \ccalH\ccalO}}}

\def \hbSo {\hbS_{{\scriptscriptstyle \ccalO}}}

\def \hbSoh {\hbS_{{\scriptscriptstyle \ccalO\ccalH}}}

\def \bbAo {\bbA_{{\scriptscriptstyle \ccalO}}}

\def \bbAoh {\bbA_{{\scriptscriptstyle \ccalO\ccalH}}}

\def \bbLo {\bbL_{{\scriptscriptstyle \ccalO}}}
\def \bbLh {\bbL_{{\scriptscriptstyle \ccalH}}}
\def \bbLoh {\bbL_{{\scriptscriptstyle \ccalO\ccalH}}}

\def \tbLo {\tbL_{{\scriptscriptstyle \ccalO}}}

\def \bbDo {\bbD_{{\scriptscriptstyle \ccalO}}}

\def \bbDoh {\bbD_{{\scriptscriptstyle \ccalO\ccalH}}}

\def \bbWo {\bbW_{{\scriptscriptstyle \ccalO}}}

\def \bbWoh {\bbW_{{\scriptscriptstyle \ccalO\ccalH}}}

\def \bbxo {\bbx_{{\scriptscriptstyle \ccalO}}}

\def \bbXo {\bbX_{{\scriptscriptstyle \ccalO}}}
\def \bbXh {\bbX_{{\scriptscriptstyle \ccalH}}}

\def \bbVo {\bbV_{{\scriptscriptstyle \ccalO}}}

\def \ccalSoh {\ccalS_{{\scriptscriptstyle \ccalO\ccalH}}}

\def \Fsco {F_{{\scriptscriptstyle \mathrm{score}}}}

\title{Learning Graphs from Smooth and Graph-Stationary Signals with Hidden Variables}
\author{Andrei Buciulea~\IEEEmembership{Student Member,~IEEE}, Samuel Rey~\IEEEmembership{Student Member,~IEEE},
        and~Antonio~G.~Marques,~\IEEEmembership{Senior Member,~IEEE}\vspace{-0.5cm}
	\thanks{Work supported by the Spanish NSF Grants SPGraph (PID2019-105032GB-I00) and FPU17/04520, by the Grants F661-MAPPING-UCI and F663-AAGNCS funded by the Comunidad de Madrid (CAM) and King Juan Carlos University
		(URJC), and by the Grants F649-1209 and PREDOC20-003 funded by the CAM and URJC. All the authors are with the Dept. of Signal Theory and Comms., King Juan Carlos University, Madrid, Spain. Email contact author: antonio.garcia.marques@urjc.es. An early preliminary version of this work was presented as a conference paper in~\cite{buciulea2019network}.
		}}
\renewcommand{\baselinestretch}{0.96} 

\begin{document}
\maketitle
\begin{abstract}
Network-topology inference from (vertex) signal observations is a prominent problem across data-science and engineering disciplines. Most existing schemes assume that observations from all nodes are available, but in many practical environments, only a subset of nodes is accessible. A natural (and sometimes effective) approach is to disregard the role of unobserved nodes, but this ignores latent network effects, deteriorating the quality of the estimated graph. Differently, this paper investigates the problem of inferring the topology of a network from nodal observations while taking into account \textit{the presence of hidden (latent) variables}.
Our schemes assume the number of observed nodes is considerably larger than the number of hidden variables and build on recent graph signal processing models to relate the signals and the underlying graph. Specifically, we go beyond classical correlation and partial correlation approaches and assume that the signals are \emph{smooth} and/or \emph{stationary} in the sought graph. The assumptions are codified into different constrained optimization problems, with the presence of hidden variables being explicitly taken into account. Since the resulting problems are ill-conditioned and non-convex, the block matrix structure of the proposed formulations is leveraged and suitable convex-regularized relaxations are presented.
Numerical experiments over synthetic and real-world datasets showcase the performance of the developed methods and compare them with existing alternatives.
\end{abstract}
\begin{IEEEkeywords}
Network-topology inference, hidden nodes, latent variables, graphical Lasso, graph stationarity.
\end{IEEEkeywords}
%

%%%%%%%%%%%%%%%%%%%%%%%%%%%%%%%%%%%%%%%%%%%%%%%%%%%%%%%%%%%%%%%%%%%%%%%%%%%%%%%%%%%%%%%%%%%%%%%%%%%%%%%%%%%%%%%%%%%%%%%%%%%
\section{Introduction}\label{S:Introduction}
%%%%%%%%%%%%%%%%%%%%%%%%%%%%%%%%%%%%%%%%%%%%%%%%%%%%%%%%%%%%%%%%%%%%%%%%%%%%%%%%%%%%%%%%%%%%%%%%%%%%%%%%%%%%%%%%%%%%%%%%%%%
% Leveraging the graph and GSP
Recent years have witnessed the rise of problems involving datasets with non-Euclidean support.
A popular approach to deal with this type of data consists in exploiting graphs to generalize a wide range of classical information-processing techniques to those irregular domains. 
This graph-based perspective has been successfully applied to a number of applications (with power, communications, social, geographical, genetics, and brain networks being notable examples~\cite{kolaczyk2009book,EmergingFieldGSP,sporns2012book,nodop1998field,rey2019sampling}) and has attracted the attention of researchers from different areas, including statistics, machine learning and signal processing (SP).
For the latter case, graph SP (GSP) has been capable of generalizing a number of tools originally conceived to process signals with regular support (time or space) to signals defined on heterogeneous domains represented by a graph, providing new insights and efficient algorithms~\cite{EmergingFieldGSP,SandryMouraSPG_TSP13,SandryMouraSPG_TSP14Freq,djuric2018cooperative,marques2020editorial}.
The core assumption of GSP is that the properties of the graph signals can be explained by the influence of the network, whose topology is codified in the so-called graph-shift operator (GSO), a square matrix whose non-zero entries identify the edges of the graph.

% Topology inference
Although networks may exist as physical entities, oftentimes they are abstract mathematical representations with nodes describing variables and links describing pairwise relationships between them. 
More importantly for the paper at hand, such relationships may not be always known a priori.
In the scenarios where the graph is unknown, it is possible to learn the graph from a set of nodal observations under the fundamental assumption that there exists a relationship between the properties of the observed signals and the topology of the sought graph.
The described task represents a prominent problem commonly referred to as \textit{network topology inference}, which is also known as \textit{graph learning}~\cite{Kalofolias2016inference_smoothAISTATS16,pavez_laplacian_inference_icassp16,segarra2017network,segarra2018network,mateos2019connecting,sardellitti2019graph}.
Noteworthy approaches include correlation networks~\cite{kolaczyk2009book}, partial correlations and (Gaussian) Markov random fields~\cite{meinshausen06,GLasso2008,kolaczyk2009book,Lake10discoveringstructure}, sparse structural equation models~\cite{BazerqueGeneNetworks,BainganaInfoNetworks}, GSP-based approaches~\cite{MeiGraphStructure,DongLaplacianLearning, pavez_laplacian_inference_icassp16, segarra2017network,segarra2018network}, as well as their non-linear generalizations \cite{Karanikolas_icassp16,shen2016kernelsTSP16}, to name a few.

% Motivating the presence of hidden variables and related works
The standard network-inference approach in the aforementioned works is to assume that observations from all the nodes of the graph are available. In certain environments, however, only observations from a subset of nodes are available, with the remaining nodes being unobserved or \textit{hidden}.
The existence of hidden/latent nodes constitutes a relevant and challenging problem since closely related values from two observed nodes may be explained not only by an edge between the two nodes but by a third latent node connected to both of them.
Moreover, because there are no observations from the hidden nodes, modeling their influence renders the network inference problem substantially more challenging and ill-posed.
Except for direct pairwise methods, which can be trivially generalized to the setup at hand, most of the existing approaches require important modifications to deal with hidden nodes. Network-inference works that have looked at the problem of hidden variables include examples in the context of Gaussian graphical model selection~\cite{chandrasekaran2012latent,yang2020network}, inference of linear Bayesian networks~\cite{anandkumar2013learning}, nonlinear regression~\cite{mei2018silvar}, or brain connectivity~\cite{chang2019graphical} to name a few. Nonetheless, there are still a number of effective network-inference methods (including most in the context of GSP) that have not considered the presence of latent unobserved nodes. 

% State our contribution: motivating smoothness and stationarity
Motivated by the previous discussion, in this paper we approach the problem of network topology inference with hidden variables by leveraging two fundamental concepts of the GSP framework: smoothness~\cite{EmergingFieldGSP} and stationarity~\cite{marques2016stationaryTSP16,perraudinstationary2016}.
A signal being smooth on a graph implies that the signal values at two neighboring nodes are close so that the signal varies slowly across the graph.
This fairly general assumption has been successfully exploited to infer the topology of the graph when values from all nodes are observed~\cite{DongLaplacianLearning,kalofolias2018large,wang2021efficient}.
From a different perspective, assuming that a random process is stationary on a graph is tantamount to assuming that the covariance matrix of the random process is a polynomial of the GSO, which has been leveraged in the context of network-inference to develop new algorithms and establish important links between graph stationarity and classical correlation and partial-correlation approaches ~\cite{segarra2017network,shafipour2020online,navarro2020joint}.
% Overview of our approach
Although the assumptions of smoothness and stationarity have been successfully adopted in the context of the network-topology inference problem, a formulation robust to the presence of hidden variables is still missing.
To fill this gap, this paper builds on our previous work and investigates how the presence of the hidden variables impacts the classical definitions of graph smoothness and stationarity.
Then, it formulates the network-recovery problem as a constrained optimization that accounts explicitly for the modified definitions.
A key in our formulation is the consideration of a block matrix factorization approach and exploitation of the low rankness and the sparsity pattern present in the blocks related to hidden variables.
A range of formulations are presented and suitable (convex and non-convex) relaxations to deal with the sparsity and low-rank terms are considered. While our focus is to learn the connections among observed nodes, some of our approaches also reveal information related to links involving hidden nodes.
A further investigation of this matter is left as future work. 

% Summary of contributions
To summarize, our contributions are as follows: (i) we analyze the influence of hidden variables on graph smoothness and graph stationarity; (ii) we propose several optimization problems to solve the topology inference problem with hidden variables when the observed signals are smooth, stationary, or both; and (iii) we present an extensive evaluation of the proposed models through both synthetic and real experiments. 

% Paper outline
The remainder of the paper is organized as follows.
Section~\ref{S:fundamental_GSP} introduces basic GSP concepts leveraged during the paper.
Section~\ref{S:hidden_variables_inference} formalizes the problem at hand.
Sections~\ref{S:smooth_inf} and~\ref{S:stationary_inf} respectively detail the proposed topology inference algorithms for smooth and stationary signals, with Section~\ref{S:smooth_stationary_inf} combining both assumptions and considering that the signals are both smooth and stationary.
The numerical evaluation of the proposed methods is presented in Section~\ref{S:numerical_experiments}, and Section~\ref{S:conclusions} provides some concluding remarks.

%%%%%%%%%%%%%%%%%%%%%%%%%%%%%%%%%%%%%%%%%%%%%%%%%%%%%%%%%%%%%%%%%%%%%%%%%%%%%%%%%%%%%%%%%%%%%%%%%%%%%%%%%%%%%%%%%%%%%%%%%%%
\section{Fundamentals of graph signal processing}\label{S:fundamental_GSP}
%%%%%%%%%%%%%%%%%%%%%%%%%%%%%%%%%%%%%%%%%%%%%%%%%%%%%%%%%%%%%%%%%%%%%%%%%%%%%%%%%%%%%%%%%%%%%%%%%%%%%%%%%%%%%%%%%%%%%%%%%%%
% Introduction to basic GSP concepts. G,S,A,L.
In this section, we introduce basic GSP concepts that help to explain the relationship between the observed signals and the topology of the underlying graph. 
    
\vspace{1mm}
\noindent\textbf{GSO and graph signals.}
Let $\ccalG = \{ \ccalV, \ccalE\}$ be an undirected and weighted graph with $N$ nodes where $\ccalV$ and $\ccalE$ represent the vertex and edge set, respectively.
The weighted adjacency matrix $\bbA\in\reals^{N \times N}$ is a sparse matrix encoding the topology of the graph $\ccalG$, with $A_{ij}$ capturing the weight of the edge between the nodes $i$ and $j$, and with $A_{ij}=0$ if $i$ and $j$ are not connected.
A general representation of the graph is the GSO $\bbS \in \mathbb{R}^{N \times N}$, where $S_{ij} \neq 0$ if and only if $i=j$ or $(i,j) \in \ccalE$.
Typical choices for the GSO are the adjacency matrix $\bbA$ \cite{SandryMouraSPG_TSP13}, the combinatorial graph Laplacian  $\bbL := \diag(\bbA \textbf{1}) - \bbA$ \cite{EmergingFieldGSP}, and their degree-normalized variants.
Since the graph is undirected, the GSO is symmetric and can be diagonalized as $\bbS=\bbV\bbLambda \bbV^\top$, where the orthogonal matrix $\bbV\in\reals^{N \times N}$ collects the eigenvectors of the GSO and the diagonal matrix $\bbLambda$ its eigenvalues.
A graph signal can be denoted as a vector $\bbx\in \mathbb{R}^{N}$ where $x_{i}$ represents the signal value observed at node $i$.
A common tool to model the relationship between the signal $\bbx$ and its underlying graph are the graph filters. 
A graph filter $\bbH\in\reals^{N\times N}$ is a linear operator defined as a polynomial of the GSO of the form
\begin{equation}\label{E:graph_filter}
    \bbH = \sum_{l=0}^{L-1} h_{l} \bbS^{l}=\bbV  \sum_{l=0}^{L-1} h_{l}\bbLambda^l\bbV^\top=\bbV\diag(\tbh)\bbV^\top,
\end{equation}
where the filter degree is $L-1$, $\{h_l\}_{l=0}^{L-1}$ represent the filter coefficients, and $\tbh\in\reals^N$ denotes the frequency response of the graph filter.
Since $\bbH$ is a polynomial of $\bbS$, it readily follows that both matrices have the same eigenvectors.

\vspace{1mm}
\noindent\textbf{Graph stationarity.} A \textit{random} graph signal $\bbx$ is stationary on the graph $\ccalG$ if it can be represented as the output of a graph filter $\bbH$ with a zero mean white signal $\bbw \in \mathbb{R}^{N}$ as input, i.e., the covariance of $\bbw$ is $\mathbb{E}[\bbw\bbw^\top]=\bbI$ and $\bbx=\bbH\bbw$.
In turn, if $\bbx$ is stationary, then its covariance $\bbC$ is given by
\begin{equation}\label{E:cov_x}
    \bbC= \mathbb{E}[\bbx\bbx^\top]=\bbH\mathbb{E}[\bbw\bbw^\top]\bbH^\top=\bbH\bbH^\top=\bbH^2.
\end{equation}
In the spectral domain, it can be seen from \eqref{E:cov_x} that the GSO $\bbS$ and the covariance matrix $\bbC$ share the same eigenvectors $\bbV$~\cite{marques2016stationaryTSP16,perraudinstationary2016,Girault15}.
Therefore, graph stationarity implies that the matrices $\bbS$ and $\bbC$ commute, i.e., $\bbC\bbS = \bbS\bbC$, which is a relevant property to be exploited later on.
    
\vspace{1mm}
\noindent\textbf{Graph smoothness.}
A graph signal is considered smooth on a graph $\ccalG$ if the signal value at two connected nodes is ``close'', or equivalently, if the difference between the signal value at neighboring nodes is small. 
A common approach to quantify the smoothness of a graph signal is by means of the quadratic form~\cite{Kalofolias2016inference_smoothAISTATS16}
\begin{equation}\label{E:smoothness}
    \sum_{(i,j)\in\ccalE}{A_{ij}(x_{i}-x_{j})^2} = \bbx^\top\bbL\bbx,
\end{equation}
which quantifies how much the signal $\bbx$ changes with respect to the notion of similarity encoded in the weights of $\bbA$.
This measure will be referred to as ``local variation'' ($\mathrm{LV}$) of $\bbx$.
Note that, if the goal is to obtain the mean $\mathrm{LV}$ of $M$ graph signals collected in the $N \times M$ matrix $\bbX=[\bbx_1,...,\bbx_M]$, this can be achieved by computing
\begin{equation}\label{E:localvariation_tracecovariance}
\frac{1}{M}\sum_{m=1}^M \bbx_m^\top\bbL\bbx_m=\frac{1}{M}\sum_{m=1}^M \tr(\bbx_m\bbx_m^\top\bbL)=\tr(\hbC\bbL), 
\end{equation} 
where $\hbC:=\frac{1}{M}\sum_{m=1}^M \bbx_m\bbx_m^\top=\frac{1}{M}\bbX\bbX^\top$ denotes the sample estimate of the covariance of $\bbX$.

%%%%%%%%%%%%%%%%%%%%%%%%%%%%%%%%%%%%%%%%%%%%%%%%%%%%%%%%%%%%%%%%%%%%%%%%%%%%%%%%%%%%%%%%%%%%%%%%%%%%%%%%%%%%%%%%%%%%%%%%%%%
%\section{Topology inference with hidden variables: problem formulation}\label{S:hidden_variables_inference}
\section{Influence of hidden variables in the topology inference model}\label{S:hidden_variables_inference}
%%%%%%%%%%%%%%%%%%%%%%%%%%%%%%%%%%%%%%%%%%%%%%%%%%%%%%%%%%%%%%%%%%%%%%%%%%%%%%%%%%%%%%%%%%%%%%%%%%%%%%%%%%%%%%%%%%%%%%%%%%%
% Intro
The current section is devoted to formally posing the topology-inference problem when only observations from a subset of nodes of the graph are available.
We present a general formulation and highlight the influence of the hidden variables.

% Notation of hidden variables
Denote as $\bbX = [\bbx_{1},...,\bbx_{M}] \in \mathbb{R}^{N \times M}$ the collection of $M$ signals defined on top of the \emph{unknown} graph $\ccalG$ with $N$ nodes.
%, and let the random generative model of $\bbX$ be associated with the graph $\ccalG$.
Then, we consider that we only observe the values of $\bbX$ from a subset of nodes $\ccalO \subset \ccalV$ with cardinality $O<N$. In contrast, the values corresponding to the remaining $H = N-O$ nodes in the subset $\ccalH=\ccalV\setminus \ccalO$ stay hidden\footnote{With a slight abuse of notation, we use $H$ to denote the number of hidden nodes and the square matrix $\bbH$ to denote a generic graph filter.}.
%Note that classical approaches assume that observations form all nodes are available, so $\ccalO = \ccalV$.
For simplicity and without loss of generality, let the observed nodes correspond to the first $O$ nodes of the graph, so the values of the given signals at $\ccalO$ are collected in the submatrix $\bbXo\in\reals^{O\times M}$, which is formed by the first $O$ rows of the matrix $\bbX$.
As explained in the previous section, these observations can be used to form the sample covariance matrix.
%When doing so, it is important to notice that the matrices $\bbS\in\reals^{N\times N}$ and $\hbC\in\reals^{N\times N}$, which respectively represent the GSO and the sample covariance matrix associated with the full graph $\ccalG$ and signals $\bbX\in\reals^{N\times M}$, present the following block structure
When doing so, it is important to notice the matrices $\bbS\in\reals^{N\times N}$ and $\hbC\in\reals^{N\times N}$, which respectively represent the GSO and the sample covariance matrix associated with the full graph $\ccalG$, and the signals $\bbX$, present the following block structure
\begin{equation}\label{E:block_SC}
    \bbX =
    \begin{bmatrix}
    \bbXo  \\
    \bbXh
    \end{bmatrix}\!,
    \;\bbS =
    \begin{bmatrix}
    \bbSo & \bbSoh  \\
    \bbSho & \bbSh
    \end{bmatrix}\!, \;
    \hbC =
    \begin{bmatrix}
    \hbCo & \hbCoh  \\
    \hbCho & \hbCh
    \end{bmatrix}. \;
\end{equation}
The $O \times O$ matrix $\bbSo$ denotes the GSO describing the connections between the observed nodes, while the remaining blocks model the edges involving hidden nodes. 
Similarly, $\hbCo=\frac{1}{M}\bbXo\bbXo^\top$ denotes the sample covariance of the observed signals, and the other blocks denote the submatrices of $\hbC$ involving signal values from the hidden nodes.
Since $\ccalG$ is undirected, both $\bbS$ and $\hbC$ are symmetric, and thus, $\bbSho=\bbSoh^\top$ and $\hbCho=\hbCoh^\top$.
%Note that the block structure of $\bbS$ and $\bbC$ will play a central role in the following sections to incorporate additional structure when modeling the effect of the hidden variables in the topology inference problem.

% Problem formulation
With the previous definitions in place, the problem of graph learning/network topology inference in the presence of hidden variables is formally introduced next.
\begin{problem}\label{general_problem}
    Let $\ccalG=(\ccalV,\ccalE)$ be a graph with $N$ nodes and GSO $\bbS\in\reals^{N\times N}$, and suppose that $\{\ccalV,\ccalE, N,\bbS\}$ are all unknown. Given the nodal subset $\ccalO\subset \ccalV$ with cardinality $|\ccalO|=O$, and the observations $\bbXo\in\reals^{O \times M}$ corresponding to the values of $M$ graph signals observed at the nodes in $\ccalO$, find the underlying graph structure encoded in $\bbSo\in\reals^{O\times O}$ under the assumptions that: \\
    (AS1) The number of hidden variables (nodes) is substantially smaller than the number of observed nodes, i.e., $ O \lessapprox N$; and \\
    (AS2) There exists a (known) property relating the full graph signals $\bbX\in\reals^{N \times M}$  to the GSO $\bbS$.
\end{problem}
Despite having observations from $O$ nodes, there are still $H=N-O$ nodes that remain unseen and influence the observed signals $\bbXo$, rendering the inference problem challenging and severely ill-conditioned.
To make the problem more tractable, (AS1) ensures that the number of hidden variables is small.
Assumption (AS2) is more generic and establishes that there is a known relationship between the graph signals $\bbX$ and the full graph $\bbS$.
The particular relationship is further developed in the following sections, where we assume that $\bbX$ is either smooth (Section~\ref{S:smooth_inf}) or stationary (Section~\ref{S:stationary_inf}) on $\bbS$.
The key issue to address is how (AS2), which involves the full signals and GSO, translates to the submatrices $\bbXo$, $\bbSo$, and $\bbCo$ in \eqref{E:block_SC}.

% General framework
Given the above considerations, a general formulation to solve Problem~\ref{general_problem} is as follows
\begin{alignat}{2}\label{E:general_inf_prob}
\!\!\hbSo=&\;\mathrm{argmin}_{\bbSo}  \
&& f(\bbSo)                                    \\ 
\!\!&\!\mathrm{\;\;\;\;\;s. \;t. } &&\bbXo \in \ccalX(\bbS),  \;\; \nonumber
\\
\!\!&\! && 
 \bbSo \in \ccalS, \;\; \nonumber
\end{alignat}
where $f(\cdot)$ is a (preferably convex) function that promotes desirable properties on the sought graph. 
Typical examples include the $\ell_{1}$ norm, the Frobenius norm, the spectral radius, or linear combinations of those \cite{mateos2019connecting}.
Note that the first constraint in \eqref{E:general_inf_prob} (referred to as observation constraint) takes into account that $\ccalX$ involves the full matrices $\bbX$ and $\bbS$ but only $\bbXo$ is observed. 
It is also important to remark that, as will be apparent in the following sections, for observations that are either smooth or stationarity in the graph, the constraint $\bbXo \in \ccalX(\bbS)$ can be reformulated in terms of the (sample) covariance matrices  $\hbCo=\frac{1}{M}\bbXo\bbXo^\top$ and $\bbCo=\E{\bbxo\bbxo^\top}$.
%Since $\bbXo$ depends on the number of samples $M$, and under the assumption that $M\gg N$, to develop more efficient algorithms we rely on $\bbC$ rather than $\bbX$. 
%Note that from the setting described in Problem~\ref{general_problem} we can always obtain the sample estimate $\hbCo$ from the signals $\bbXo$.
%Then, the set $\ccalC(\bbS)$ captures the constraints that model the relationship between $\bbX$ and $\bbS$ stated in (AS2), guaranteeing that the estimated graph present some desired properties and reducing the size of the feasible set. 
%When problem \eqref{E:general_inf_prob} is particularized in the following sections these constraints may be incorporated to  the objective function.
Regarding the second constraint in \eqref{E:general_inf_prob}, the set $\ccalS$ collects the requirements for $\bbS$ to be a specific type of GSO.
A typical example is the set of adjacency matrices
\begin{equation}\label{E:A_set}
\ccalA := \{ A_{ij} \geq 0; \ \bbA = \bbA^\top; \ A_{ii} = 0; \ \bbA\mathbf{1}\geq\mathbf{1}\},
\end{equation}
where we require the GSO to have non-negative weights, be symmetric, and have no self-loops, and the last constraint rules out the trivial $0$ solution by imposing that every node has at least one neighbor. Analogously, the set of combinatorial Laplacian matrices is
\begin{equation}\label{E:L_set}
\ccalL := \{ L_{ij} \leq 0 \;\mathrm{for} \; i \neq j; \; \bbL=\bbL^\top; \;   \bbL \textbf{1} = \bb0; \; \bbL  \succeq 0 \},
\end{equation}
where we require the GSO to be a positive semidefinite matrix, have non-positive off-diagonal values, have positive entries on its diagonal, and have the constant vector as an eigenvector (i.e, the sum of the entries of each row to be zero). Lastly, we want to stress that the objective $f(\bbSo)$ and the constraint $\bbSo \in \ccalS$ can be alternatively formulated based on the full GSO $\bbS$, provided that we know that the structural properties (for instance sparsity in the objective and positive entries in the constraints) hold also for the non-observed parts of $\bbS$. Such an approach is suitable when the interest goes beyond $\bbSo$ and spans the estimation of the links involving the nodes in $\ccalH$.

% Other works
\vspace{.2cm}
\noindent \textbf{Hidden variables in correlation and partial-correlation networks:} Before discussing our specific solutions to Problem~\ref{general_problem}, a relevant question is how classical topology-inference approaches (namely correlation and partial-correlation networks) handle the problem of latent nodal variables.
The so-called direct methods consider that a link between nodes $i$ and $j$ exists based only on a pairwise similarity metric between the signals observed at $i$ and $j$. Within this class of methods, correlation networks set the similarity metric to the correlation and, as a result, $\bbS$ corresponds to a (thresholded) version of $\bbC$.
Given their simplicity, the generalization of direct methods to setups where hidden variables are present is straightforward and simply given by $\bbSo = \hbCo$.
Nevertheless, a high correlation between two nodes can be due to global network effects rather than to the direct influence among pairs of neighbors, calling for more involved topology-inference methods. To that end, partial-correlation methods, including the celebrated graphical Lasso (GL) algorithm \cite{kolaczyk2009book}, propose estimating the graph as a matrix of partial correlation coefficients, which boils down to assuming that the connectivity patterns can be identified as $\bbS=\bbC^{-1}$, with $\bbC^{-1}$ being known as the precision matrix. When hidden variables are present, the submatrix of the precision matrix is given by $\bbCo^{-1} = \bbSo - \bbB$, with $\bbB = \bbSoh\bbSh^{-1}\bbSho$ being a low-rank matrix since $H \ll O$.
Leveraging this structure, the authors in \cite{chandrasekaran2012latent} modified the GL algorithm to deal with hidden variables via a maximum-likelihood estimator augmented with a nuclear-norm regularizer to promote low rankness in $\bbB$.
The resulting algorithm is known as latent variable graphical Lasso (LVGL) and is given by
\begin{align}\label{E:def_LGL}
    \begin{split}
    \underset{\bbSo-\bbB \succeq\mathbf{0}, \, \bbB\succeq\mathbf{0}}{\max} \,\log\det(\bbSo - \bbB) -\textrm{trace}(\hbCo (\bbSo - \bbB))\\
    \hspace{3.3cm}-\lambda_1\|\bbSo\|_1  -\lambda_2\|\bbB\|_*,
    \end{split}
\end{align}
where $\hbCo$ represents the sample covariance of the observed data and $\lambda_1$ and $\lambda_2$ are regularization constants \cite{chandrasekaran2012latent}.

% Link with next section
Rather than assuming that the relation between $\bbX$ and $\bbS$ postulated in (AS2) is given by either correlations or partial-correlations, this paper looks at setups where the operational assumption is that the observed signals are: i) smooth on the graph; ii) stationary on the graph; and iii) both smooth and stationary. Sections \ref{S:smooth_inf}-\ref{S:smooth_stationary_inf} deal with each of those three setups. Section~\ref{S:numerical_experiments} evaluates numerically the performance of the developed algorithms and compares it with that of classical correlation and LVGL schemes.

%%%%%%%%%%%%%%%%%%%%%%%%%%%%%%%%%%%%%%%%%%%%%%%%%%%%%%%%%%%%%%%%%%%%%%%%%%%%%%%%%%%%%%%%%%%%%%%%%%%%%%%%%%%%%%%%%%%%%%%%%%%
\section{Topology inference from smooth signals}\label{S:smooth_inf}
%%%%%%%%%%%%%%%%%%%%%%%%%%%%%%%%%%%%%%%%%%%%%%%%%%%%%%%%%%%%%%%%%%%%%%%%%%%%%%%%%%%%%%%%%%%%%%%%%%%%%%%%%%%%%%%%%%%%%%%%%%%
% Intro explaining we adopt smoothness assumptions
In this section, we address Problem~\ref{general_problem} by particularizing  \eqref{E:general_inf_prob} to the case of the signals $\bbX$ being smooth on $\ccalG$. 

% Formulation of smoothness measure accounting for hidden variables
As explained in Section \ref{S:fundamental_GSP}, a natural way of measuring the smoothness of (a set of) graph signals is to leverage the graph Laplacian and compute their $\mathrm{LV}$ as $\frac{1}{M}\tr(\bbX\bbX^\top\bbL)$ [cf. \eqref{E:localvariation_tracecovariance}]. As a result, in this section we set $\bbS=\bbL$ and focus on $\hbC=\frac{1}{M}\bbX\bbX^\top$. Recall that, due to the existence of hidden variables, the whole covariance matrix is not observed. To account for this and leveraging the block definition of $\hbC$ and $\bbS$ introduced in \eqref{E:block_SC}, we can rewrite the $\mathrm{LV}$ of our dataset as
\begin{equation}\label{E:smooth_C_hidd}
    \tr(\hbC\bbL) = \tr(\hbCo\bbLo) + 2\tr(\hbCoh\bbLoh^\top) + \tr( \hbCh\bbLh),
\end{equation}
where only $\hbCo=\frac{1}{M}\bbXo\bbXo^\top$ is assumed to be known and the influence of the hidden variables in the  $\mathrm{LV}$ has been made explicit.
%Note that if $\bbCo$ is not perfectly known, it can be trivially replaced by its sampled estimate $\hbCo$.

% Leveraging structure
Although the block-wise smoothness presented in \eqref{E:smooth_C_hidd} could be directly employed to approach the network-topology inference as an optimization problem, most of the submatrices are not known and need to be estimated. Incorporating the terms $\bbCoh\bbLoh^\top$ and $\bbCh\bbLh$ would directly render the problem non-convex.
To circumvent this issue, we lift the problem by defining the matrix $\bbK:=\bbCoh\bbLoh^\top\in\reals^{O \times O}$. Since (AS1) guarantees that $\rank(\bbK)\leq H \ll O$, we exploit the low-rank structure of the matrix $\bbK$ in our formulation.
Correspondingly, we also define the matrix $\bbR:=\bbCh\bbLh\in\reals^{H \times H}$ and note that, since $\bbR$ is the product of two positive semidefinite matrices, it has positive eigenvalues and, as a result, it holds that $\tr(\bbR)\geq 0$.

% First formulation - Problem with Lo
With these considerations in mind, the network topology inference from smooth signals is formulated as
\begin{alignat}{2}\label{E:smooth_hidden_Lo}
    \!\!&\!\min_{\bbLo,\bbK,\bbR} \
    && \tr(\bbCo\bbLo) \! + \! 2\tr(\bbK) \! + \! \tr(\bbR) \! + \!\alpha \| \bbLo \|_{F,off}^{2} \! \nonumber \\
    \!\!&\! && \;\;\;\; - \beta \log(\diag(\bbLo)) + \gamma \| \bbK \|_{*}                                    \\ 
    \!\!&\!\mathrm{\;\;s. \;t. } && \;\;\;\;\tr(\bbCo\bbLo) + 2\tr(\bbK) + \tr(\bbR) \geq 0, \;\; \nonumber \\
    \!\!&\! &&  \;\;\;\; \tr(\bbR) \geq 0, \;\; \nonumber \\
    \!\!&\! &&  \;\;\;\; \bbLo \in \bar{\ccalL}, \;\; \nonumber
\end{alignat}
where $\|\cdot\|_{F,off}^2$ denotes the Frobenius norm excluding the elements of the diagonal. This term, together with $\log(\diag(\bbLo))$, serves to control the sparsity of $\bbLo$.
Furthermore, the logarithmic barrier rules out the trivial solution of $\bbLo=\mathbf{0}$.
The nuclear norm $\|\cdot\|_*$ is a convex regularizer that promotes low-rank solutions for the matrix $\bbK$ and it is typically employed as a surrogate of the (non-convex) rank constraint.
%The adoption of the nuclear norm, together with the consideration of the matrices $\bbK$ and $\bbR$, ensure the convexity of \eqref{E:smooth_hidden_Lo}, and hence, a globally optimum solution to \eqref{E:smooth_hidden_Lo} can be efficiently found.
The adoption of the nuclear norm, together with the consideration of the matrices $\bbK$ and $\bbR$, ensure the convexity of \eqref{E:smooth_hidden_Lo} so a globally optimum solution can be efficiently found.
The weights $\alpha,\beta,\gamma \geq 0$ control the trade-off between the regularizers, the first constraint ensures that the LV is non-negative, and the second constraint captures that fact of matrix\footnote{From an algorithmic point of view, it is worth noticing that the matrix $\bbR$ always appears as $\tr(\bbR)$ in \eqref{E:smooth_hidden_Lo}. As a result, if convenient to reduce the numerical burden, one can replace $\tr(\bbR)$ with $r$ and optimize over $r$ in lieu of $\bbR$. See the related formulation in \eqref{E:smooth_hidden_Lao} for details.} $\bbR$ being PSD.  

The last point to discuss in detail is the form of $\bar{\ccalL}$. 
Mathematically, the set $\bar{\ccalL}$ is equivalent to the set of combinatorial Laplacians $\ccalL$, but replacing the condition $\bbL\mathbf{1}=\mathbf0$ with $\bbL \textbf{1} \geq \bb0$, i.e., $\bar{\ccalL} :=  \{L_{ij} \leq 0 \; \mathrm{for} \; i \neq j; \; \bbL=\bbL^\top; \; \bbL \textbf{1} \geq \bb0; \; \bbL  \succeq 0\}$.
The modification is required because, strictly speaking, $\bbLo$ is not a combinatorial Laplacian. The existence of links between the elements in $\ccalO$ and the hidden nodes in $\ccalH$ give rise to non-zero (negative) entries in $\bbLoh$ and, as a result, the sum of the off-diagonal elements of $\bbLo$ can be smaller than the value of the associated diagonal elements (which account for the links in both $\ccalO$ and $\ccalH$). 
Intuitively, the more relaxed condition $\bbLo\mathbf{1}\geq \mathbf{0}$ enlarges the set of feasible solutions rendering the inference process harder to solve, an issue that has been observed when running the numerical experiments.
Moreover, when estimating the diagonal of $\bbLo$ we are indirectly estimating the number of edges between observed and the hidden nodes. This could be potentially leveraged to estimate links with non-observed nodes, but this entails a more challenging problem that goes beyond the scope of the paper. An approach to bypass some of these issues is analyzed next.

\subsection{Exploiting the Laplacian of the observed adjacency matrix}\label{S:L_ao}
% Motivation Laplacian of Ao
The Laplacian $\bbL$ offers a neat way to measure the smoothness of graph signals [cf. \eqref{E:smoothness}]. However,  when addressing the problem of estimating the Laplacian from smooth signals under the presence of hidden nodes, we must face the challenges associated with the fact of the submatrix $\bbLo$ not being a Laplacian itself. As discussed in the preceding paragraphs, this requires dropping some of the Laplacian constraints from the optimization, leading to a looser recovery framework. 
To circumvent these issues, rather than estimating $\bbLo$, this section looks at the problem of estimating $\tbLo:=\diag(\bbAo\mathbf{1}) -\bbAo$, the Laplacian associated with the observed adjacency matrix $\bbAo \in \reals^{O \times O}$. In contrast to $\bbLo$, the matrix $\tbLo$ is a proper combinatorial Laplacian ($\tbLo \in \ccalL$) and, hence, the original Laplacian constraints can be restored. 
%Then, it is pertinent to analyze the effect of employing the matrix $\tbLo$ instead of $\bbLo$ when solving \eqref{E:smooth_hidden_Lo}. 
The remaining of this section is devoted to reformulating \eqref{E:smooth_hidden_Lo} in terms of $\tbLo$.

% Analyzing influence of Laplacian Ao
Upon defining the $O \times O$ diagonal matrices $\bbDo:=\diag(\bbAo\mathbf{1})$ and $\bbDoh:=\diag(\bbAoh\mathbf{1})$, which count the number of observed and hidden neighbors for the nodes in $\ccalO$, the matrix $\bbLo$ is expressed as $\bbLo = \bbDo + \bbDoh - \bbAo = \tbLo + \bbDoh$.
With this equivalence, the smoothness penalty in \eqref{E:smooth_hidden_Lo} is rewritten as
\begin{align}
    \tr(\bbC\bbL) &= \tr(\bbCo\tbLo) + \tr(\bbCo\bbDoh) + 2\tr(\bbK)  + \tr(\bbR) \nonumber \\
    &= \tr(\bbCo\tbLo) + 2\tr(\tbK)  + \tr(\bbR),
\end{align}
where $\tbK:=\bbCo\bbDoh/2+\bbK$.
Because the entries of $\bbDoh$ depend on the presence of edges between the observed and the hidden nodes, if the graph is sparse, the matrix $\bbDoh$ will be a low-rank matrix.
Furthermore, since the sparsity pattern of the diagonal of $\bbDoh$ depends on the matrix $\bbAoh=-\bbLoh$, it follows that the column sparsity pattern of $\bbCo\bbDoh$ matches that of $\bbK$, and thus, $\tbK$ is also low rank.

With these considerations in mind, we reformulate the optimization in \eqref{E:smooth_hidden_Lo} replacing $\bbLo$ with $\tbLo$, resulting in the following convex optimization problem
\begin{alignat}{2}\label{E:smooth_hidden_Lao}
    \!\!&\!\min_{\tbLo,\tbK,r} \
    && \tr(\bbCo\tbLo) \! + \! 2\tr(\tbK) \! + \! r \! + \!\alpha \| \tbLo \|_{F,off}^{2} \! \\
    \!\!&\! && \;\;\;\; - \beta \log(\diag(\tbLo)) + \! \gamma_{*} \| \tbK \|_{*}  \! + \! \gamma_{2,1} \| \tbK \|_{2,1}  \nonumber                                   \\ 
    \!\!&\!\mathrm{\;\;s. \;t. } && \;\;\;\; \tr(\bbCo\tbLo) + 2\tr(\tbK) + r \geq 0, \;\; \nonumber \\
    \!\!&\!  && \;\;\;\;r  \geq 0 \nonumber \\
    \!\!&\! &&  \;\;\;\; \tbLo \in \ccalL, \;\; \nonumber
\end{alignat}
where $\bar{\ccalL}$ in \eqref{E:smooth_hidden_Lo} has been replaced with $\ccalL$ in \eqref{E:smooth_hidden_Lao}, which is the set of all valid combinatorial Laplacian matrices defined in \eqref{E:L_set}. 
%Note that, although we have also replaced $\bbK$ with $\tbK$, where the terms associated with $\bbK$ in the objective and costraints of \eqref{E:smooth_hidden_Lo} are the same than those associated with $\tbK$ in \eqref{E:smooth_hidden_Lao}. 
%Hence, the rank of $\tbK$ depends on the sparsity of the graph, which must be taken into account when choosing the weights $\gamma_*$ and $\gamma_{2,1}$. Lastly, knowing that $\tr(\bbR) \geq 0$ we optimize over a nonnegative variable $r$ instead of matrix $\bbR$.
Moreover, knowing that the matrix $\bbR$ only appears as $\tr(\bbR)$ we replace it with the nonnegative variable $r$ to alleviate the numerical burden.
Note that, although we replaced $\bbK$ with $\tbK$, the terms previously associated with $\bbK$ in \eqref{E:smooth_hidden_Lo} remain unchanged in \eqref{E:smooth_hidden_Lao}.
Nonetheless, while the original matrix $\bbK\in\reals^{O\times O}$ is low rank because it is the product of a tall $O \times H$ matrix and a fat $H \times O$ matrix, the low-rankness of $\tbK$ is a byproduct of the sparsity of the graph.
More precisely, the matrix $\tbK$ involves the product of the square (full rank) matrix $\bbCo$ and the diagonal matrix $\bbDoh$.
Since the diagonal of $\bbDoh$ is sparse, such a product gives rise to a matrix with several zero columns, with the rank of the resultant matrix coinciding with the number of non-zero columns.
We exploit this structure by further regularizing  the matrix $\tbK$ with the $\ell_{2,1}$ norm.
%The original matrix $\bbK\in\reals^{O\times O}$ is low rank because it is the product of a tall $O \times H$ matrix and a fat $H \times O$ matrix.
%In contrast, the matrix $\tbK$ involves the product of the square (full rank) matrix $\bbCo$ and the diagonal matrix $\bbDoh$. Since the diagonal of $\bbDoh$ is sparse, such a product gives rise to a matrix with several zero columns, with the rank of the resultant matrix coinciding with the number of non-zero columns. 

%Note that two different configurations of the algorithm in \eqref{E:smooth_hidden_Lao} can be obtained depending on the values of the regularizers.
Indeed, two different configurations of \eqref{E:smooth_hidden_Lao} can be obtained depending on the values of the regularization constants.
Setting $\gamma_{2,1}=0$ we promote a solution with a low rank on $\tbK$ by applying the nuclear norm regularization.
Since the nuclear norm minimization does not ensure the desired column-sparsity of $\tbK$, an alternative is to set $\gamma_*=0$ and rely on the penalty $\|\tbK\|_{2,1}$. 
The computation of $\|\tbK\|_{2,1}$ can be understood as a two-step process where one first obtains the $\ell_2$ norm of each of the columns of $\tbK$ and, then, the $\ell_1$ norm of the resulting row vector is computed.
This regularization is commonly known as the group Lasso penalty~\cite{yuan2006model,simon2013sparse} and has been used in a number of sparse-recovery problems. 
The results in Section~\ref{S:numerical_experiments} will illustrate that the formulation in \eqref{E:smooth_hidden_Lao} succeeds in promoting the desired column-sparsity pattern when using the appropriate values for the hyperparameters $\gamma_*$ and $\gamma_{2,1}$. 
Note also that, by looking at the non-zero columns of $\tbK$, the nodes in $\ccalO$ with connections to hidden nodes can be identified.

%%%%%%%%%%%%%%%%%%%%%%%%%%%%%%%%%%%%%%%%%%%%%%%%%%%%%%%%%%%%%%%%%%%%%%%%%%%%%%%%%%%%%%%%%%%%%%%%%%%%%%%%%%%%%%%%%%%%%%%%%%%
\section{Topology inference from stationarity signals}\label{S:stationary_inf}
%%%%%%%%%%%%%%%%%%%%%%%%%%%%%%%%%%%%%%%%%%%%%%%%%%%%%%%%%%%%%%%%%%%%%%%%%%%%%%%%%%%%%%%%%%%%%%%%%%%%%%%%%%%%%%%%%%%%%%%%%%%
% Intro for stationary prior
In this section, instead of relying on the smoothness of the signals $\bbX$, we approach Problem~\ref{general_problem} by modifying (AS2) and considering that the data is stationary on the sought graph.  
% Hidden variables and stationary signals
The assumption of $\bbX$ being stationary on $\ccalG$ is tantamount to the matrices $\bbC$ and $\bbS$ sharing the same eigenvectors $\bbV$~\cite{marques2016stationaryTSP16}. As a result, the approach for the fully observable case is to use the observations to estimate the sample covariance $\hbC$ and then rely on the sample covariance to estimate the eigenvectors $\bbV$~\cite{segarra2017network}. However, when dealing with hidden variables, there is no obvious way to obtain $\bbVo$, the submatrix of the eigenvectors of the full covariance, using as input the submatrix $\hbCo$. To bypass this problem, instead of requiring the eigenvectors of $\bbC$ and $\bbS$ being the same, our approach is to require that $\bbC$ and $\bbS$ commute, i.e., that the equation $\bbC\bbS=\bbS\bbC$ must hold~\cite{segarra2017joint}.   
To see why this condition leads to a more tractable formulation, let us leverage the block structure of $\bbC$ and $\bbS$ described in \eqref{E:block_SC}. It follows readily that the upper left submatrix of size $O \times O$ in both sides of the equality $\bbC\bbS=\bbS\bbC$ is given by
\begin{equation}\label{E:block_commutation}
    \bbCo\bbSo+\bbCoh\bbSoh^\top=\bbSo\bbCo+\bbSoh\bbCoh^\top.
\end{equation}
The above expression succeeds in relating the sought $\bbSo$ with $\bbCo$, which can be efficiently estimated using $\bbXo$.
Furthermore,~\eqref{E:block_commutation} reveals that when hidden variables are present, we cannot simply ask $\bbSo$ and $\bbCo$ to commute, but we also need to account for the associated terms $\bbCoh\bbSoh^\top$ and $\bbSoh\bbCoh^\top$.

% Leveraging structure and initial formulation
Implementing steps similar to those in Section~\ref{S:smooth_inf}, we can lift the problem defining the matrix $\bbK=\bbCoh\bbSoh^\top \in \reals^{O \times O}$ and leverage the fact that $\rank(\bbK)\leq H \ll O$, due to (AS1).
Note that the matrix $\bbK$ is equivalent to the one defined in Section~\ref{S:smooth_inf} with the only difference that now we use a block from the generic GSO $\bbSoh$ instead of the Laplacian $\bbLoh$.
Moreover, since both $\bbC$ and $\bbS$ are symmetric matrices, we have that $\bbK^\top=\bbSoh\bbCoh^\top$. 
Then, under the general assumption that graphs are typically sparse, we can approach Problem~\ref{general_problem} with stationary observations by solving
\begin{alignat}{2}\label{E:eqn_zero_norm}
    \!\!&\!\min_{\bbSo, \bbK} \
    &&\|\bbSo\|_0                                    \\ 
    \!\!&\!\mathrm{\;\;s. \;t. } && \;\;\;\;\bbCo\bbSo+\bbK = \bbSo\bbCo+\bbK^\top, \;\; \nonumber
    \\
    \!\!&\! && 
    \;\;\;\; \rank(\bbK)\leq H, \;\; \nonumber
    \\
    \!\!&\! && 
    \;\;\;\; \bbSo \in \ccalS, \;\; \nonumber
\end{alignat}
where the $\ell_0$ norm promotes sparse solutions, the equality constraint ensures commutativity of the GSO and the covariance while accounting for latent nodes, and the rank constraint captures the low rank of $\bbK$ due to (AS1).

Regarding the specific choice of the GSO, when the interest is in the Laplacian matrix we set $\bbSo=\tbLo$, with $\tbLo$ denoting the Laplacian of the observed adjacency matrix.
Then, the matrix $\bbK$ is replaced with $\tbK=\bbCo\bbDoh+\bbK$, which accounts for the fact of using $\tbLo$ instead of $\bbLo$ in \eqref{E:block_commutation}.
This was further motivated in Section~\ref{S:L_ao}, and the discussion provided there also applies here.

The presence of the rank constraint and the $\ell_{0}$ norm renders \eqref{E:eqn_zero_norm} non-convex and computationally hard to solve.
Furthermore, the first constraint assumes perfect knowledge of $\bbCo$, which may not always represent a practical setup.
These issues are addressed in the next section.

\subsection{Convex and robust stationary topology inference}
% Convex relaxations
A natural approach to deal with \eqref{E:eqn_zero_norm} is to relax the non-convex terms, replacing the $\ell_0$ norm with the $\ell_1$ norm and the rank constraint with the nuclear norm, their closest convex surrogates.
Furthermore, in most practical scenarios the ensemble covariance $\bbCo$ is not known and one must rely on its sampled counterpart $\hbCo$.
This requires relaxing the equality constraint $\bbCo\bbSo+\bbK = \bbSo\bbCo+\bbK^\top$ and replacing it with a constraint that guarantees that the terms on the left-hand side and right-hand side are similar but not necessarily the same.
Taking all these considerations into account, the relaxed convex topology-inference problem is 
\begin{alignat}{2}\label{E:eqn_robust_norm} 
    \!\!&\!\min_{\bbSo, \bbK} \
    &&\|\bbSo\|_1 + \eta\|\bbK\|_*                                    \\ 
    \!\!&\!\mathrm{\;\;s. \;t. } && \|\hbCo\bbSo+\bbK-\bbSo\hbCo-\bbK^\top\|_F^2\leq\epsilon, \nonumber \\
    \!\!&\!  && \bbSo \in \ccalS, \nonumber
\end{alignat} 
where $\eta \geq 0$ controls the low rankness of $\bbK$. 
Regarding the (relaxed) stationarity constraint, the squared Frobenius norm has been adopted to measure the similarity between the matrices at hand, but other (convex) distances could be alternatively used. 
It is also important to note that the value of the non-negative constant $\epsilon$ should be selected based on prior knowledge on the noise level present in the observations and, more importantly, the number of samples $M$ used to estimate the covariance.
Clearly, if $M<O$, the matrix is not full rank, increasing notably the size of the feasible set. On the other hand, if $M\rightarrow \infty$, one can set $\epsilon=0$. 
This reduces drastically the degrees of freedom of the formulation and, as a result, renders more likely the solution to \eqref{E:eqn_robust_norm}  to coincide with the actual GSO.

% Reweighted algorithm
\textit{Remark 1 (Reweighted algorithm)}: The formulation in \eqref{E:eqn_robust_norm}  is convex and robust.
However, while replacing the original $\ell_0$ norm with the convex $\ell_1$ norm constitutes a common approach, it is well-known that non-convex surrogates can lead to sparser solutions. Indeed, a more sophisticated alternative in the context of sparse recovery is to define $\delta$ as a small positive number and replace the $\ell_0$ norm with a (non-convex) logarithmic penalty $ \|\bbSo\|_0 \approx \sum_{i,j=1}^{O}\log(|[\bbSo ]_{ij}|+\delta)$ \cite{candes2008enhancing}. An efficient way to handle the non-convexity of the logarithmic penalty is to rely on a majorization-minimization (MM) approach~\cite{sun2016majorization}, which considers an iterative linear approximation to the concave objective and leads to an \emph{iterative} re-weighted $\ell_1$ minimization. To be specific, with  $t=1,...,T$ being the iteration index, adopting such an approach for the problem in \eqref{E:eqn_robust_norm}  results in  
\begin{alignat}{2}\label{E:eqn_one_norm_rew}
    \!\!&\!\bbSo^{(t+1)} := \argmin_{\bbSo, \bbK} \
    && \sum_{i,j=1}^O[\bbW^{(t)}]_{ij}|[\bbSo]_{ij}| + \eta\|\bbK\|_*                                    \\ 
    \!\!&\!\mathrm{\;\;s. \;t. } && \bbCo\bbSo+\bbK = \bbSo\bbCo+\bbK^\top, \nonumber \\
     \!\!&\!  && \bbSo \in \ccalS, \nonumber
\end{alignat}
with $\bbW^{(t)}$ being defined as $[\bbW^{(t)}]_{ij} = \left(\Big|\big[\bbSo^{(t-1)}\big]_{ij}\Big|+\delta\right)^{-1}$.
%, and $\mathrm{vec}(\cdot)$ denoting the operation that reshapes a matrix into a vector. 
Since the iterative algorithm penalizes (assigns a larger weight to) entries of $\bbSo$ that are close to zero, the obtained solution is typically sparser at the expense of a higher computational cost. Finally, note that the absolute values can be removed whenever the constraint $[\bbSo ]_{ij}\geq 0$ is enforced.

\subsection{Exploiting structure through alternating optimization}
% Column sparse structure
In the previous section, the product of the unknown matrices $\bbCoh$ and $\bbSoh^\top$ was absorbed into matrix $\bbK$. Since such a matrix is low rank, the convex nuclear norm was used to promote low-rank solutions while achieving convexity. However, when implementing this approach, there were other properties (such as $\bbSoh$ being sparse) that were ignored. A reasonable question is, hence, if the judicious incorporation of the additional information outperforms the potential loss of convexity. In this section, we propose an efficient alternating \emph{non-convex} algorithm that accounts for the additional structure present in our setup.  Its associated recovery performance (along with comparisons to its convex counterparts) will be tested in Section \ref{S:numerical_experiments}.

A well-established approach in low-rank optimization is to factorize the matrix of interest as the product of a tall and fat matrix, which boils down to replacing $\bbK$ with the original submatrices $\bbCoh$ and $\bbSoh^\top$.
Moreover, when the value of $H$ is unknown, which determines the size of $\bbCoh$ and $\bbSoh^\top$, a principled approach is to rely on an upper bound on $H$ and add the Frobenius terms $\|\bbCoh\|_F$ and $\|\bbSoh\|_F$ to the objective function (see, e.g., \cite{srebro2005rank} for a formal derivation of this approach).
In our particular setup, this factorization has the additional benefit of $\bbSoh$ being sparse.
Then, the resulting non-convex optimization problem is given by
\begin{alignat}{2}\label{E:eqn_factorized} 
    \!\!&\!\min_{\bbSo, \bbCoh, \bbSoh} \
    && \sum_{i,j=1}^{O}\log(|[\bbSo ]_{ij}|+\delta) + \eta \|\bbSoh\|_F^2 \nonumber \\ 
    \!\!&\! && +\nu\sum_{i,j=1}^{O,H}\log(|[\bbSoh]_{ij}|+\delta) + \eta \|\bbCoh\|_F^2 \nonumber \\
    \!\!&\!  && +\rho\|\hbCo\bbSo\!+\!\bbCoh\bbSoh^\top\!-\!\bbSo\hbCo\!-\!\bbSoh\bbCoh^\top\|_F^2 \nonumber \\
    \!\!&\!\mathrm{\;\;s. \;t. } && \bbSo \in \ccalS, \;\; \bbSoh \in \ccalSoh,
\end{alignat} 
Clearly, problem \eqref{E:eqn_factorized} guarantees that the rank of the matrix $\bbSoh\bbCoh^\top$ is upper bounded by the size of its composing factors $\bbSoh$ and $\bbCoh$.
In this case, the sparse solutions for $\bbSo$ and $\bbSoh$ are promoted by means of the (concave) logarithmic penalty, introduced on Remark~1.
The robust commutativity constraint is placed on the objective function as a penalty term, and the set $\ccalSoh$ captures the fact that $\bbSoh$ is a block from the GSO.
In its simplest form, we have that $\ccalSoh:=\{S_{ij}\geq0\}$ if the GSO is the adjacency matrix, and $\ccalSoh:=\{S_{ij}\leq0\}$ if it is set to the Laplacian matrix.

The main drawback associated with the formulation in \eqref{E:eqn_factorized} is that the presence of the bilinear term $\bbCoh\bbSoh^\top$ and logarithmic penalty render the problem non-convex.
To address this issue, we implement a Block Successive Upper bound Minimization (BSUM) algorithm~\cite{blockalternating}, an iterative approach that merges the techniques from MM and alternating optimization.
Then, we find a solution to \eqref{E:eqn_factorized} by iterating between the following thee steps. 

\vspace{.5mm}
\noindent \textbf{Step 1.}
Given the estimates $\hbCoh^{(t)}$ and $\hbSoh^{(t)}$, we substitute $\bbCoh=\hbCoh^{(t)}$ and $\bbSoh=\hbSoh^{(t)}$ into \eqref{E:eqn_factorized} and solve it to estimate $\bbSo$.
This yields
\begin{alignat}{2}\label{E:factorized_step1} 
    \!\!&\!\hbSo^{(t+1)} := && \argmin_{\bbSo\in\ccalS} \
    \sum_{i,j=1}^O[\bbWo^{(t)}]_{ij}|[\bbSo]_{ij}| \; \\ 
    \!\!&\! && \!\!\! +\rho\|\hbCo\bbSo\!+\!\hbCoh^{(t)}[\hbSoh^{(t)}]^\top\!-\!\bbSo\hbCo\!-\!\hbSoh^{(t)}[\hbCoh^{(t)}]^\top\|_F^2, \nonumber
\end{alignat} 
where the logarithmic penalty is approximated by the re-weighted $\ell_1$ norm as detailed after \eqref{E:eqn_one_norm_rew}.

\vspace{.5mm}
\noindent \textbf{Step 2.} Given the estimate $\hbCoh^{(t)}$ from the previous iteration, and leveraging the estimate $\hbSo^{(t+1)}$ from the last step, we estimate the matrix $\bbSoh$ by solving
\begin{alignat}{2}\label{E:factorized_step2} 
    \!\!&\!\hbSoh^{(t+1)} \!\! && := \!\!\argmin_{\bbSoh\in\ccalSoh} \
    \sum_{i,j=1}^{O,H}[\bbWoh^{(t)}]_{ij}|[\bbSoh]_{ij}| + \eta\|\bbSoh\|_F^2 \; \\ 
    \!\!&\! &&  \! +\rho\|\hbCo\hbSo^{(t+1)}\!+\!\hbCoh^{(t)}\bbSoh^\top\!-\!\hbSo^{(t+1)}\hbCo\!-\!\bbSoh[\hbCoh^{(t)}]^\top\|_F^2. \nonumber
\end{alignat} 

\vspace{.5mm}
\noindent \textbf{Step 3.} With the estimates from the previous steps in place, the last step involves estimating the matrix $\bbCoh$ by solving
\begin{alignat}{2}\label{E:factorized_step3} 
    \!\!&\!\hbCoh^{(t+1)} && :=  \argmin_{\bbCoh} \;\;\; \eta\|\bbCoh\|_F^2\\
    \!\!&\! &&  \!\!\! \|\hbCo\hbSo^{(t+1)}\!+\!\bbCoh[\hbSoh^{(t+1)}]^\top\!-\!\hbSo^{(t+1)}\hbCo\!-\!\hbSoh^{(t+1)}\bbCoh^\top\|_F^2. \nonumber
\end{alignat} 
The alternating algorithm is initialized upon: i) solving \eqref{E:eqn_robust_norm} to obtain $\hbK$ and ii) setting $\hbCoh$ and $\hbSoh$ as the $H$ top $\hbK$ left and right singular vectors of $\hbK$.
The three steps proposed in \eqref{E:factorized_step1}-\eqref{E:factorized_step3} are iterated until convergence to a stationary point, a result that is formally stated next.

\vspace{1mm}
\noindent\textbf{Proposition 1.} \textit{Denote with $f$ the objective function in \eqref{E:eqn_factorized}. 
Let $\ccalY^*$ be the set of stationary points of \eqref{E:eqn_factorized}, and let $\bby^{(t)}=[\vvec(\bbSo^{(t)})^\top\!,\vvec(\bbSoh^{(t)})^\top\!,\vvec(\bbCoh^{(t)})^\top]^\top$ be the solution generated after running the 3 steps in \eqref{E:factorized_step1}-\eqref{E:factorized_step3} $t$ times.
Then, the solution generated by the iterative algorithm \eqref{E:factorized_step1}-\eqref{E:factorized_step3} converges to a stationary point of $f$ as $t$ goes to infinity, i.e., 
}
\[\lim_{t\to\infty} d(\bby^{(t)},\ccalY^*) = 0,\]
\textit{with $d(\bby,\ccalY^*) := \min_{\bby^* \in \ccalY^*} \|\bby-\bby^*\|_2$.}
%\overset{\scriptsyle{\Delta}}{=} 

\vspace{1mm}
\noindent Note that convergence was not obvious since at least one of the steps does not have a unique minimizer, and the first and second steps employ an approximation of the objective function in \eqref{E:eqn_factorized}.
The details of the proof, which relies on convergence results for BSUM schemes \cite[Th. 1b]{blockalternating}, are provided in Appendix A.

Although more computationally expensive, the numerical tests in  Section~\ref{S:numerical_experiments} confirm that the additional structure incorporated by replacing $\bbK$ with $\bbSoh$ and $\bbCoh$ together with the re-weighted $\ell_1$ approach for encouraging sparsity give rise to a better network reconstruction, provided that the iterative optimization is initialized with the solution to the convex formulation in \eqref{E:eqn_robust_norm}.
Last but not least, notice that an additional benefit of the formulation in \eqref{E:eqn_factorized} is that, by analyzing $\hbSoh$, information of the potential links between nodes in $\ccalO$ and the hidden nodes in $\ccalH$ is obtained. While network-tomography schemes \cite{kolaczyk2009book} go beyond the scope of this paper, the results in this section can be used as a first step towards that goal.

%eqref{E:factorized_step1}-\eqref{E:factorized_step3} is stated in the next result. 

% Other works
\vspace{.2cm}
\noindent \textbf{Graph stationary vis-\`a-vis graph smoothness:}  Suppose that we are given two datasets $\bbXo$ and $\bbXo'$, both with the same number of signals. Moreover, suppose that we also know that the observed signals $\bbXo$ are smooth on an unknown graph, that $\bbXo'$ are stationary on an unknown graph, and that our goal is to identify the underlying graphs. Based on that information, we run the algorithms in Section \ref{S:smooth_inf} for the dataset $\bbXo$ and those in this section for the dataset $\bbXo'$. An interesting question is which one yields a better recovery result. While the exact answer depends on all the particularities of each of the setups, from a general point of view stationary schemes are expected to achieve better results. The reason is that stationarity strongly limits the degrees of freedom of the GSO, while smoothness is a more lenient assumption, an intuition that will be validated in Section \ref{S:numerical_experiments}. Equally relevant, there can be situations where the data is both stationary and smooth. That is the case, for example, if the covariance matrix shares the eigenvectors with the graph Laplacian and its power spectral density is low pass. In such a setup, one could combine both network-recovery approaches, leading to a better recovery performance. This is precisely the subject of the ensuing section.

%%%%%%%%%%%%%%%%%%%%%%%%%%%%%%%%%%%%%%%%%%%%%%%%%%%%%%%%%%%%%%%%%%%%%%%%%%%%%%%%%%%%%%%%%%%%%%%%%%%%%%%%%%%%%%%%%%%%%%%%%%%
\section{Topology inference from stationary and smooth graph signals with hidden variables}\label{S:smooth_stationary_inf}
%%%%%%%%%%%%%%%%%%%%%%%%%%%%%%%%%%%%%%%%%%%%%%%%%%%%%%%%%%%%%%%%%%%%%%%%%%%%%%%%%%%%%%%%%%%%%%%%%%%%%%%%%%%%%%%%%%%%%%%%%%%
% Smooth formulation with nuclear norm
In this section, we address Problem~\ref{general_problem} by assuming that the graph signals $\bbX$ are both smooth and stationary on the unknown graph $\ccalG$.
These two assumptions can be jointly considered to design optimization problems with additional structure to enhance the estimation of $\bbSo$.
To that end, we consider the smoothness-based inference problem described in \eqref{E:smooth_hidden_Lao} and incorporate the robust commutativity constraint accounting for stationarity [cf. \eqref{E:block_commutation}], resulting in
\begin{alignat}{2}\label{E:smooth_hidden_st_Lao}
    \!\!&\!\min_{\tbLo,\tbK,r} \
    && \tr(\hbCo\tbLo) \!+\! 2\tr(\tbK) + r +\!\alpha \| \tbLo \|_{F,off}^{2} \\ 
    \!\!&\! && \;\;\;\; - \beta \log(\diag(\tbLo)) \! + \! \gamma_{*} \| \tbK \|_{*}  \! + \! \gamma_{2,1} \| \tbK \|_{2,1}  \nonumber                                   \\ 
    \!\!&\!\mathrm{\;\;s. \;t. } && \tr(\hbCo\tbLo) + 2\tr(\tbK) + r \geq 0, \;\; \nonumber \\
    \!\!&\! && \tbLo \in \ccalL, \;\; \nonumber \\
    \!\!&\! &&  \| \hbCo\tbLo + \tbK -\tbLo\hbCo - \tbK^\top  \|_{F}^2 \leq \epsilon. \;\; \nonumber
\end{alignat}
Since the smooth formulation involves the Laplacian matrix, note that we adopted the Laplacian of the observed adjacency matrix as the GSO.
Regarding the stationarity constraint, as discussed for \eqref{E:eqn_robust_norm}, the value of $\epsilon$ should be selected based on the number of available signals $M$ and the observation noise. It is also worth noting that the matrix $\tbK$ is inconspicuously absorbing the error derived from the presence of the hidden variables and from using $\tbLo$ instead of $\bbLo$ in both the smoothness penalty and the commutativity constraint. Regarding matrix $\tbK$, two different regularizers are considered: the nuclear norm (to promote solutions with a low rank) and the $\ell_{2,1}$ norm (to promote column sparsity). Since having solutions with columns that are zero also reduces the rank, it is prudent to tune the value of the hyperparameters $\gamma_{*}$ and $\gamma_{2,1}$ jointly, so that the (joint) dependence between the rank and the column sparsity is kept under control.

We close the section by noting that the formulation in \eqref{E:smooth_hidden_st_Lao} is convex so that its globally optimal solution can be found efficiently. However, non-convex versions of \eqref{E:smooth_hidden_st_Lao} that leverage the re-weighted $\ell_{2,1}$ norm to promote column sparsity and factorization approaches for the low-rank penalty (similar to those used in Section \ref{S:stationary_inf}) could be developed here as well. 

% Intro next section
%In the following sections we introduce a wide gamut of numerical results to illustrate the performance of the different algorithms introduced in this paper and provide further intuition. 

%%%%%%%%%%%%%%%%%%%%%%%%%%%%%%%%%%%%%%%%%%%%%%%%%%%%%%%%%%%%%%%%%%%%%%%%%%%%%%%%%%%%%%%%%%%%%%%%%%%%%%%%%%%%%%%%%%%%%%%%%%%
\section{Numerical experiments}\label{S:numerical_experiments}
%%%%%%%%%%%%%%%%%%%%%%%%%%%%%%%%%%%%%%%%%%%%%%%%%%%%%%%%%%%%%%%%%%%%%%%%%%%%%%%%%%%%%%%%%%%%%%%%%%%%%%%%%%%%%%%%%%%%%%%%%%%

%
\begin{figure*}
	\centering
	\begin{minipage}[c]{.33\textwidth} %.33 .27
		\includegraphics[width=\textwidth]{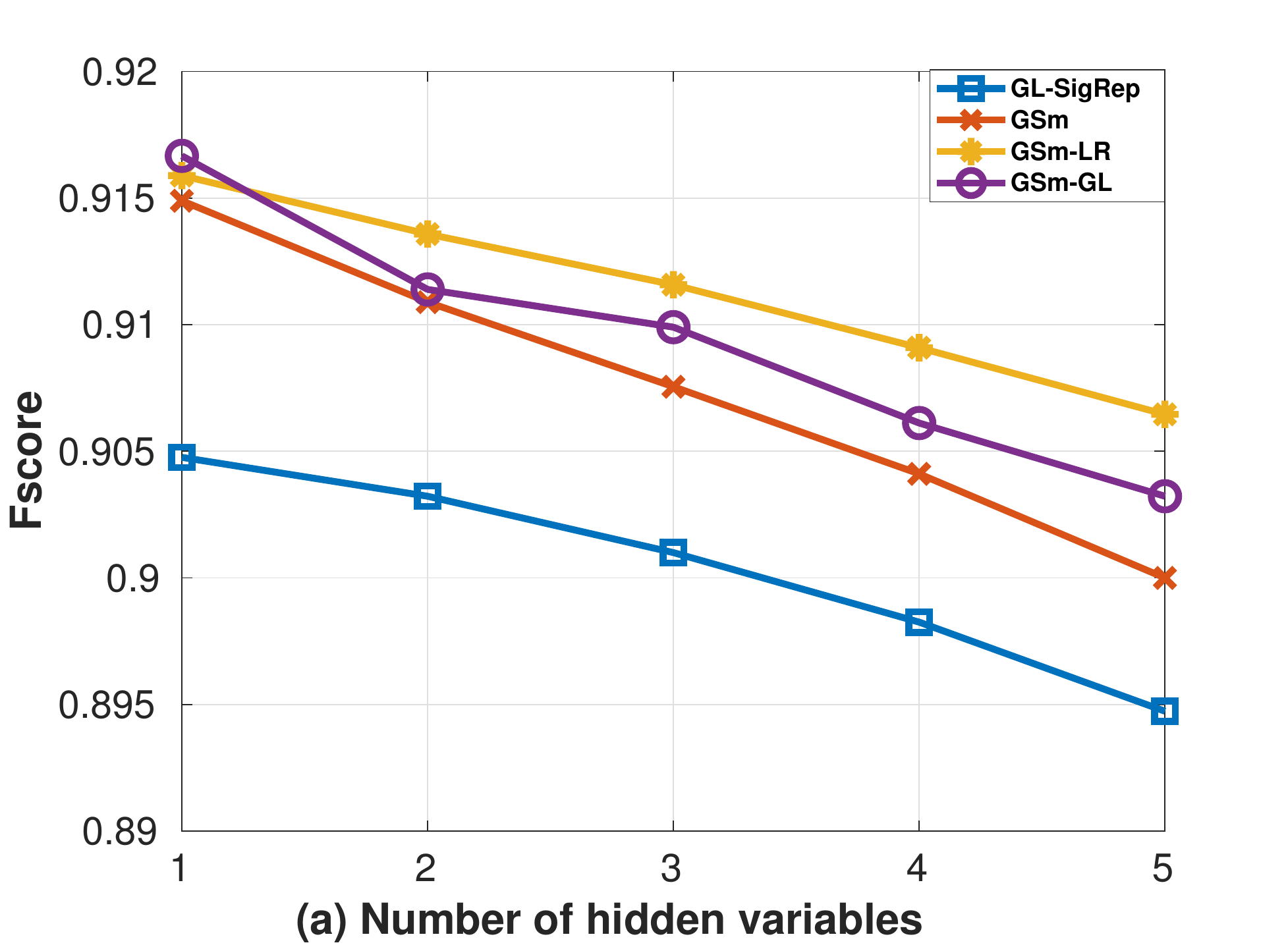}
		%\centering{\small (a)}
		\label{F:exp1}
	\end{minipage}%
	\begin{minipage}[c]{.33\textwidth} % .32 .27
		\includegraphics[width=\textwidth]{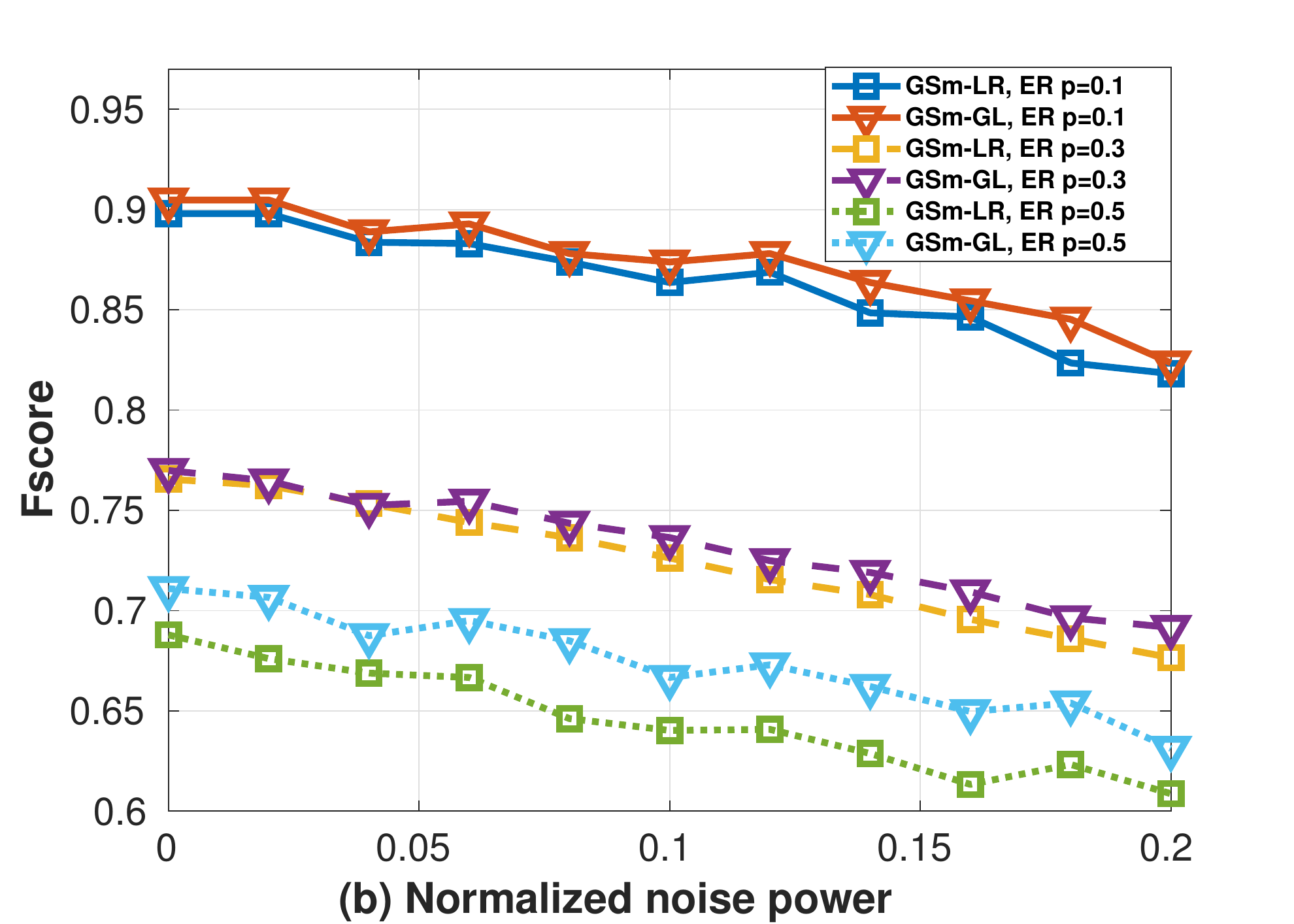}
		%\centering{\small (b)}
		\label{F:exp2}
	\end{minipage}%
	\begin{minipage}{.33\textwidth} % .16 .21
		\includegraphics[width=\textwidth]{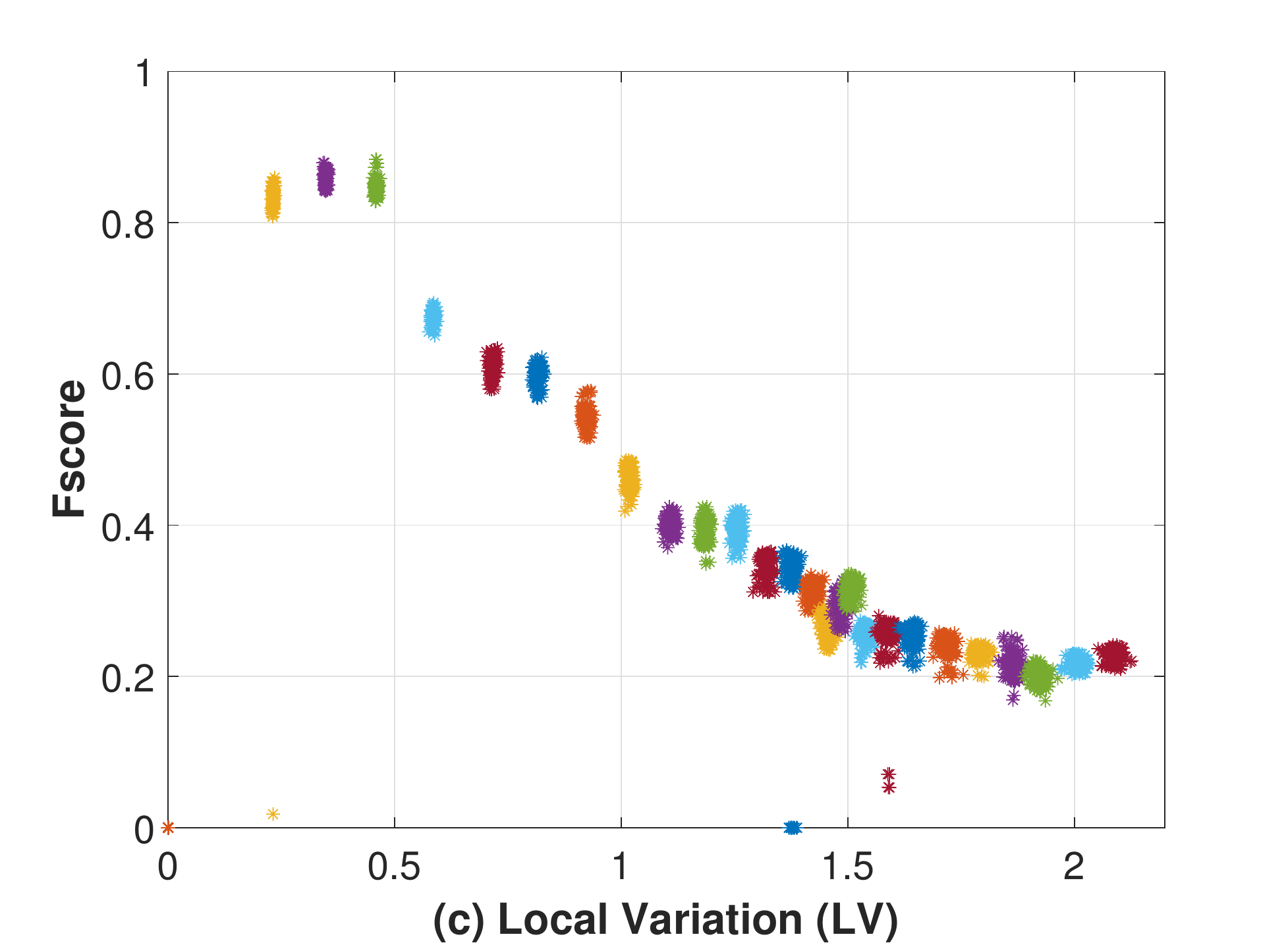}
		%\centering{\small (c)}
		\label{F:exp3}
	\end{minipage}
	\caption{Median $\Fsco$ for the algorithms based on smooth graph signals with $N = 20$ and $M=100$. The different panels assess the impact of varying  (a) the number of hidden variables $H$ for different algorithms when using RBF graphs, (b) the noise level present in the observations $\bbX$ when using Erd\H{o}s Rényi graphs with different link probabilities $p=\{0.1,0.3,0.5\}$, and (c) the average level of $LV$ of the observations $\bbX$ for a GSm-LR algorithm when  using RBF graphs.}
	%exp a: 1024 grafos 
	%exp b: 128 grafos
	%exp c: 128 realizaciones por cada señal
	%\red{AGM: Hay que cambiar el caption.}}
	%\red{AGM: qué es based on smooth? "basado en suave" ni en castellano suena bien.}
	
	\label{F:exp123}
\end{figure*}
%

%\sre{No diría que está dividida en 3 subsecciones, como mucho que tiene varias subsecciones.}
This section runs numerical experiments to gain insights on the proposed schemes and evaluate their recovery performance.
First, we test the smooth-based approaches with synthetic data and compare the results with existent algorithms from the literature.
Secondly, we assess the performance of the stationary-based schemes proposed in Section \ref{S:stationary_inf}, comparing them with those in Section \ref{S:smooth_stationary_inf} and the classical LVGL.
Lastly, we apply the proposed algorithms to two real-world datasets and compare the obtained results with those of existing alternatives.\footnote{ The MATLAB scripts for running all the numerical experiments presented in this section as well as additional related test cases can be found in \url{https://github.com/andreibuciulea/topoIDhidden}}

\subsection{Synthetic experiments based on smooth signals}\label{Ss:NumExperimentsSmoothSynthetic}
% General setting
We start by defining the default setup for the experiments in this section. With $\bbL=\bbV \bbLambda\bbV^\top$ denoting the eigendecomposition of the graph Laplacian, the smooth signals $\bbX$ are generated as $\bbX = \bbV \bbZ$, where the columns of $\bbZ\in\reals^{N\times M}$ are independent realizations of a multivariate Gaussian distribution $\bbZ \sim \mathcal{N}(\bb0,\,\bbLambda^{\dagger})$. Note that this model, which is oftentimes referred to as factor analysis \cite{factoranalysis1994,factoranalysis2011,DongLaplacianLearning}, assigns more energy to the low-frequency components, promoting smoothness on the generated graph signals. Unless otherwise stated, the number of signals is set to $M=100$ and the number of nodes to $N=20$. 
Moreover, to measure the recovery performance of the algorithms, in this section we focus on unweighted graphs and employ the $\Fsco$, which is defined as
\begin{equation}
    \Fsco = 2 \cdot \frac{precision \cdot recall}{precision + recall},
\end{equation}
where $precision$ indicates the percentage of estimated edges that are edges of the ground-truth graph and $recall$ the percentage of existing edges that were correctly estimated.

\vspace{0.2cm}
\noindent \textbf{Influence of hidden nodes.}
The results in Figure~\ref{F:exp123}.a show the variation of the $\Fsco$, as the number of hidden variables $H$ increases, for different recovery algorithms.
Graphs are randomly generated using the model in~\cite{DongLaplacianLearning}, where nodes are placed in the unit square uniformly at random and edges are computed with a Gaussian radial basis function (RBF) as $A_{ij}=\exp(-d(i,j)^2/2\sigma^2)$, with $d(i,j)$ being the euclidean distance between two vertices and $\sigma=0.5$.
Edges with weights smaller than $0.75$ are removed and the surviving ones are set to 1.
The hidden nodes are chosen uniformly at random among all the nodes in the graph.
The algorithms considered in this experiment are the following: 
(i) GL-SigRep refers to the algorithm presented in \cite{DongLaplacianLearning};
(ii) GSm is a modified version of GL-SigRep that incorporates the logarithmic penalty and relies on the sample covariance matrix $\hbC$ for the smoothness term in the objective function;
(iii) GSm-LR represents the low-rank regularized algorithm proposed in \eqref{E:smooth_hidden_Lao}, with $\gamma_{2,1}=0$; 
and (iv) GSm-GL denotes the algorithm described in \eqref{E:smooth_hidden_Lao}, with $\gamma_{*}=0$, where column-sparsity is promoted in $\tbK$ via group Lasso.
Comparing GL-SigRep with  GSm allows us to quantify the improvement obtained exclusively from including hidden variables in the formulation, providing a fairer analysis of the proposed algorithms.
The results in Figure~\ref{F:exp123}.a indicate that, although the performance of all the algorithms deteriorates when the number of hidden variables increases, the algorithms GSm-LR and GSm-GL that account for the presence of hidden variables, outperform the alternatives.
Moreover, their performance drops slowly as $H$ increases, demonstrating the importance of taking into account the presence of hidden variables.
The overall decay was expected since a higher number of hidden variables renders the topology inference problem more challenging and ill-posed, confirming the importance of (AS1).
Comparing GSm-LR with GSm-GL, we observe that their performance is similar since the generated graphs are sufficiently sparse.
It is also worth mentioning that the GSm scheme clearly outperforms GL-SigRep, illustrating the benefits of replacing the formulation introduced in~\cite{DongLaplacianLearning} with the one presented in this paper, which relies on the matrix $\hbC$ and the logarithmic barrier. 

\vspace{0.2cm}
\noindent \textbf{Noisy smooth observations.}
The second experiment assumes that the observations $\bbXo$ correspond to the ground-truth signals corrupted by additive white Gaussian noise (AWGN). For that setup, we evaluate the link-identification performance upon evaluating the $\Fsco$ achieved by GSm-LR and GSm-GL schemes, as the power of the AWGN increases, for graphs with different sparsity levels.
In the experiments, we use Erd\H{o}s Rényi (ER) graphs with edge probability values of $p = \{0.1,0.3, 0.5\}$ and set the number of hidden variables to $ H=1$.
The results, shown in  Figure~\ref{F:exp123}.b, reveal that the performance of the algorithms deteriorates not only when the noise increases but also for higher values of $p$. This behavior is consistent with the discussion provided in Section~\ref{S:smooth_inf}, since the formulation assumes that sparsity exists and, as a result, promotes solutions where several of the columns of $\tbK$ are zero.
Furthermore, we observe that GSm-LR and GSm-GL have similar performance for lower values of $p$, but when the graphs become denser GSm-GL outperforms GSm-LR.
This illustrates the fact that the low-rank regularization $\|\tbK\|_{*}$ is more sensitive to the sparsity of the graph than the group Lasso penalty $\|\tbK\|_{2,1}$.
It is also worth noting that, even though the proposed schemes were not designed to specifically account for noisy observations, the rate at which $\Fsco$ decays is smaller than the rate at which the noise power increases, showcasing the ``natural'' robustness to noise of the proposed schemes.

\begin{figure*}
	\centering
	
	\begin{minipage}[c]{.99\textwidth} %.33 .27
		\includegraphics[width=\textwidth]{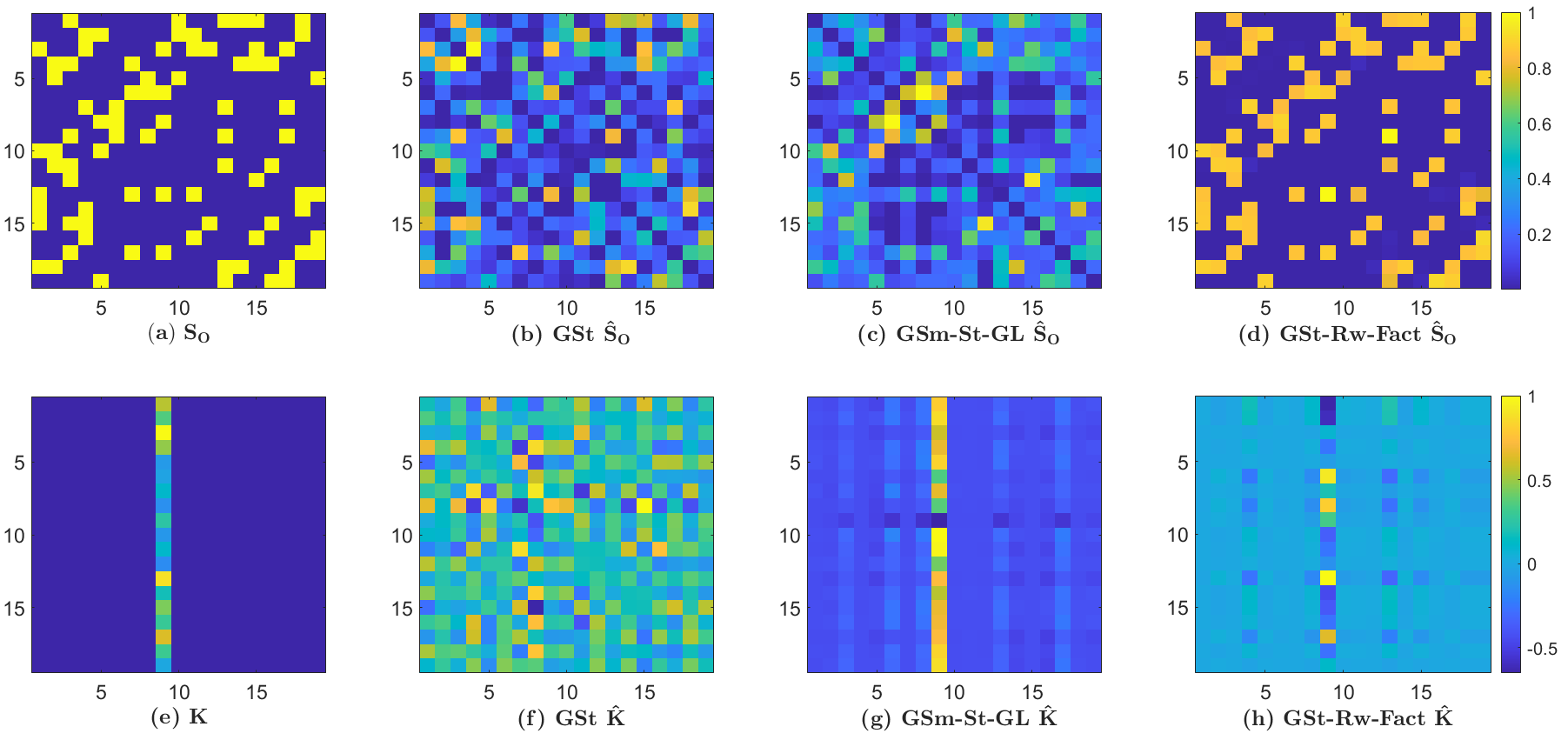}
		%\centering{\small (a)}
	\end{minipage}%
	\caption{Graphical representation of the estimates of matrices $\bbSo$ (top row) and $\bbK=\bbCoh\bbSoh^\top$ (bottom row) for different algorithms that assume the observed signals to be stationary on the graph, with $N=20$ and $H=1$. The ground-truth matrices $\bbSo$ and $\bbK$ are represented in the first column [cf. panels (a) and (e)]. Analogously, the estimates $\hbSo$ and $\hbK$ generated by GSt are represented in panels (b) and (f), those generated by GSm-St-GL in panels (c) and (g), and those generated by GSt-Rw-Fact in (d) and (h).}
	
	\label{F:exp4}
\end{figure*}

\vspace{0.2cm}
\noindent \textbf{Influence of the LV level.}
Next, we assess the relevance of the smoothness prior to the performance of the GSm-LR scheme. 
To that end, Figure~\ref{F:exp123}.c depicts the $\Fsco$ obtained with this scheme for different values of LV.
Note that as we move to the right on the $x$-axis, the observed signals exhibit a larger variation (higher frequency) and, as a result, are less smooth.
To control the LV level, the signals are generated combining $K$ successive eigenvectors as $\bbX = \bbV_K\bbZ$, with $\bbV \in \mathbb{R}^{O \times K}$ and $\bbZ \sim \mathcal{N}(\bb0,\,\bbI) \in \mathbb{R}^{K \times M}$.
The smoothest signals are obtained by selecting the first $K$ eigenvectors since they are associated with the low-frequency components. In contrast, activating the $K$ last eigenvectors maximizes the local variation of the graph signals.
For this experiment, we set $H=1$, $K=5$, and $N=30$. The first generated signal is associated with eigenvectors $k=1,...,5$, the second one with eigenvectors $k=2,...,6$, and the last ($26$th) one with eigenvectors $k=26,...,30$.
The link-identification performance for those 26 types of signals are shown in Figure~\ref{F:exp123}.c, where the vertical axis represents the $\Fsco$ and the horizontal axis the average LV $\tr(\bbX^\top\bbL\bbX)/M$.
Each color represents a different set of active frequencies and, for each set, $128$ realizations of $\bbZ$ have been generated (corresponding to the cloud of points shown in the figure). The results highlight the importance %of the smoothness assumption
of the low values of LV when assuming smooth signals on the graph since the link identification performance decays noticeably as the signal becomes high-pass.

\subsection{Synthetic experiments based on stationary signals}
In these experiments, we focus on signals that are stationary on the sought GSO $\bbS$. To facilitate comparisons with GL, two different signal models are considered: (i) $\bbC_{poly}$ and (ii) $\bbC_{MRF}$. 
For the first one, the covariance of the observed signals is generated as a random polynomial of the GSO of the form $\bbC_{poly}=\bbH^2$ with $\bbH=\sum_{l=0}^L h_l \bbS^l$, where $h_l$ are random coefficients following a normalized zero-mean Gaussian distribution. 
Note that this generative model guarantees that the covariance is PSD and a polynomial (of degree $2L$) of the GSO. 
In the second model, the covariance is generated as $\bbC_{MRF}=(\sigma\bbI+\delta\bbS)^{-1}$, where $\sigma$ is some positive number large enough to guarantee that $\bbC_{MRF}^{-1}$ is PSD and $\delta$ is some positive random number. 
As in the previous case, this generation guarantees the covariance matrix to be PSD and a polynomial of the GSO. 
Moreover, it also guarantees that the sparsity pattern of $\bbC_{MRF}^{-1}$ coincides with that of the GSO $\bbS$, which is the model assumed by GL.
Regarding the metric used to evaluate the performance, rather than using the $\Fsco$, we will generate multiple graphs and report the ratio of graphs that have been perfectly recovered (i.e., those graphs for which  \emph{all} the entries of the associated $\bbSo$ are estimated correctly). 
The reason for using this metric is that the incorporation of the stationary constraints boosts the ability of the algorithm to identify the topology, so that the value of $\Fsco$ will be very close to one for all tested schemes, rendering the comparison more difficult. 
Differently, reporting the ratio of graphs perfectly recovered helps us to better assess the differences between the tested algorithms. 

\vspace{0.2cm}
\noindent \textbf{Leveraging the structure of $\bbK$.}
While the ultimate goal of this work is to recover $\bbSo$, the properties of matrix $\bbK$ played a key role in developing several of our topology-inference algorithms. For that reason, the goal of this experiment is to illustrate the recovered (estimated) $\hbSo$ and $\hbK$, so that we can gain insights on the effectiveness of the different approaches considered in the manuscript and their influence in recovering the graph.
The results are shown in Figure~\ref{F:exp4}, where the first row represents the GSOs and the second row the matrices $\bbK$.
The first column corresponds to the ground-truth values, and the second, third and fourth columns present the estimates obtained with the low-rank scheme GSt [cf. \eqref{E:eqn_robust_norm}], the group Lasso scheme GSm-St-GL [cf. \eqref{E:smooth_hidden_st_Lao} with $\gamma_{*}=0$], and the factorized scheme GSt-Rw-Fact [cf. \eqref{E:factorized_step1}-\eqref{E:factorized_step3}], respectively.
First, focusing on $\hbK$, it is apparent that for the depicted example the low-rank scheme GSt is not capable of recovering the column-sparse structure of the original matrix $\bbK$.
Differently, when using either the group Lasso regularization (Figure~\ref{F:exp4}.g) or the factorized approach (Figure~\ref{F:exp4}.h), the estimated $\hbK$ exhibits a row-sparsity pattern that is close to that of the ground truth.
More importantly, when looking at the estimated $\hbSo$ we observe that, as desired, the more accurate estimation of $\bbK$ translates into a superior estimation of the  network topology, with GSm-St-GL yielding better estimates than GSt and GSt-Rw-Fact outperforming GSm-St-GL due to the replacement of the $\ell_1$ norm with the linearized version of the logarithmic penalty.
Overall, we believe that this simple experiment provides further intuition and strengthens the discussion about the different regularizers presented in Sections~\ref{S:smooth_inf} and~\ref{S:stationary_inf}. The next step is to test the stationary-based schemes in a more systematic way, which is the goal of the following subsections.

\begin{figure*}
	\centering
	
	\begin{minipage}[c]{.33\textwidth} %.33 .27
		\includegraphics[width=\textwidth]{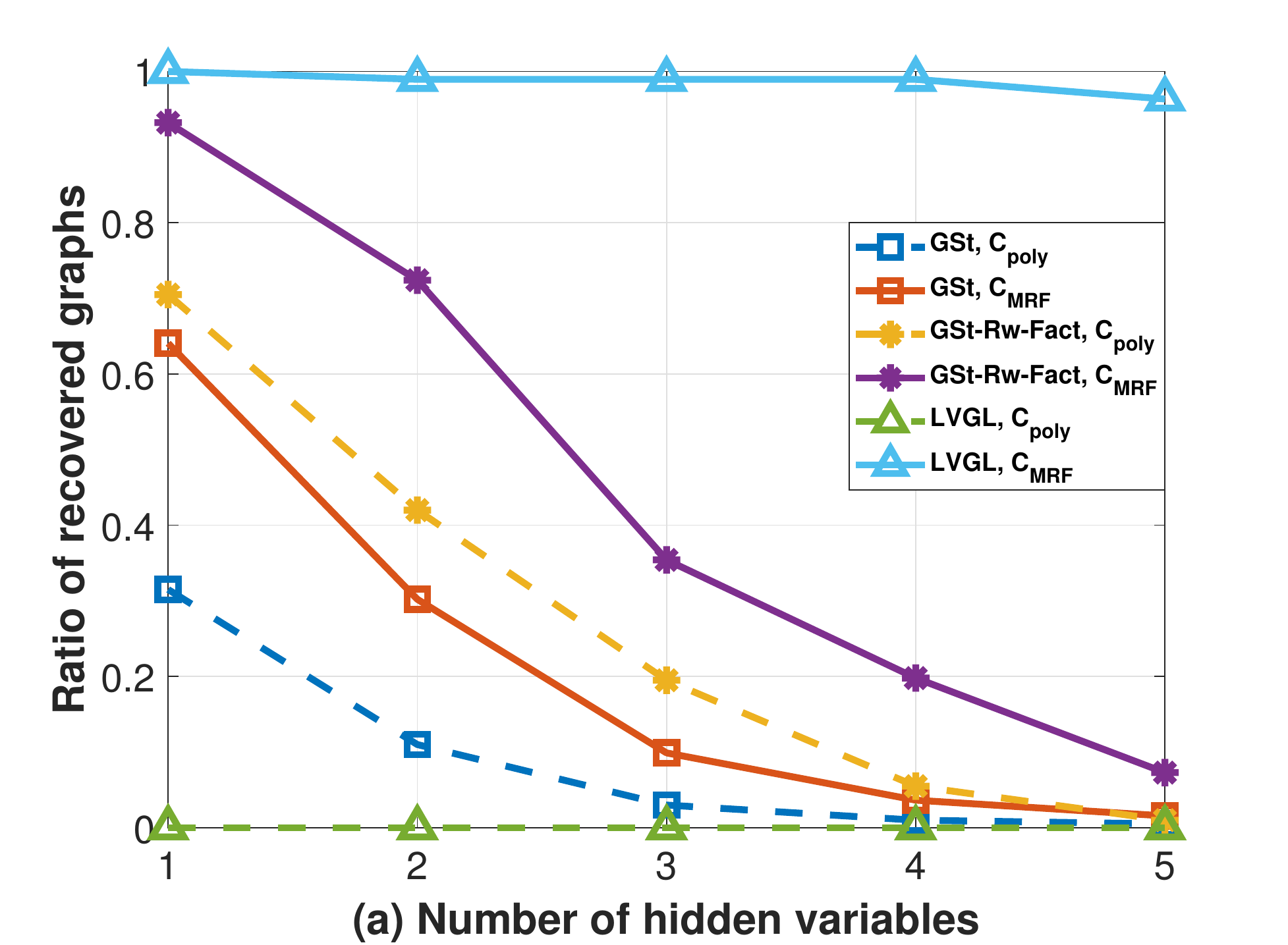}
		%\centering{\small (a)}
	\end{minipage}%
	\begin{minipage}[c]{.33\textwidth} % .32 .27
		\includegraphics[width=\textwidth]{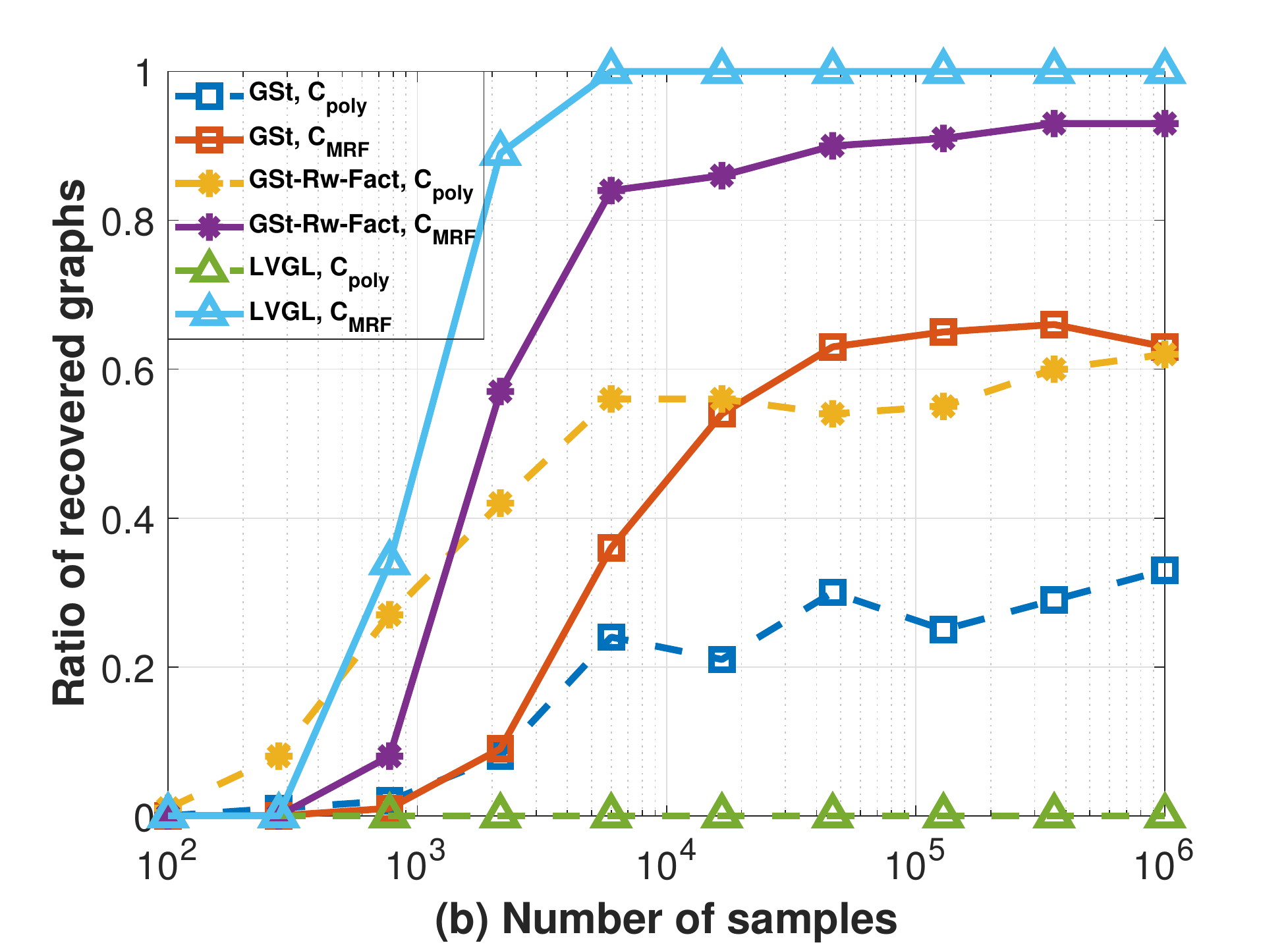}
		%\centering{\small (b)}
	\end{minipage}%
	\begin{minipage}{.33\textwidth} % .16 .21
		\includegraphics[width=\textwidth]{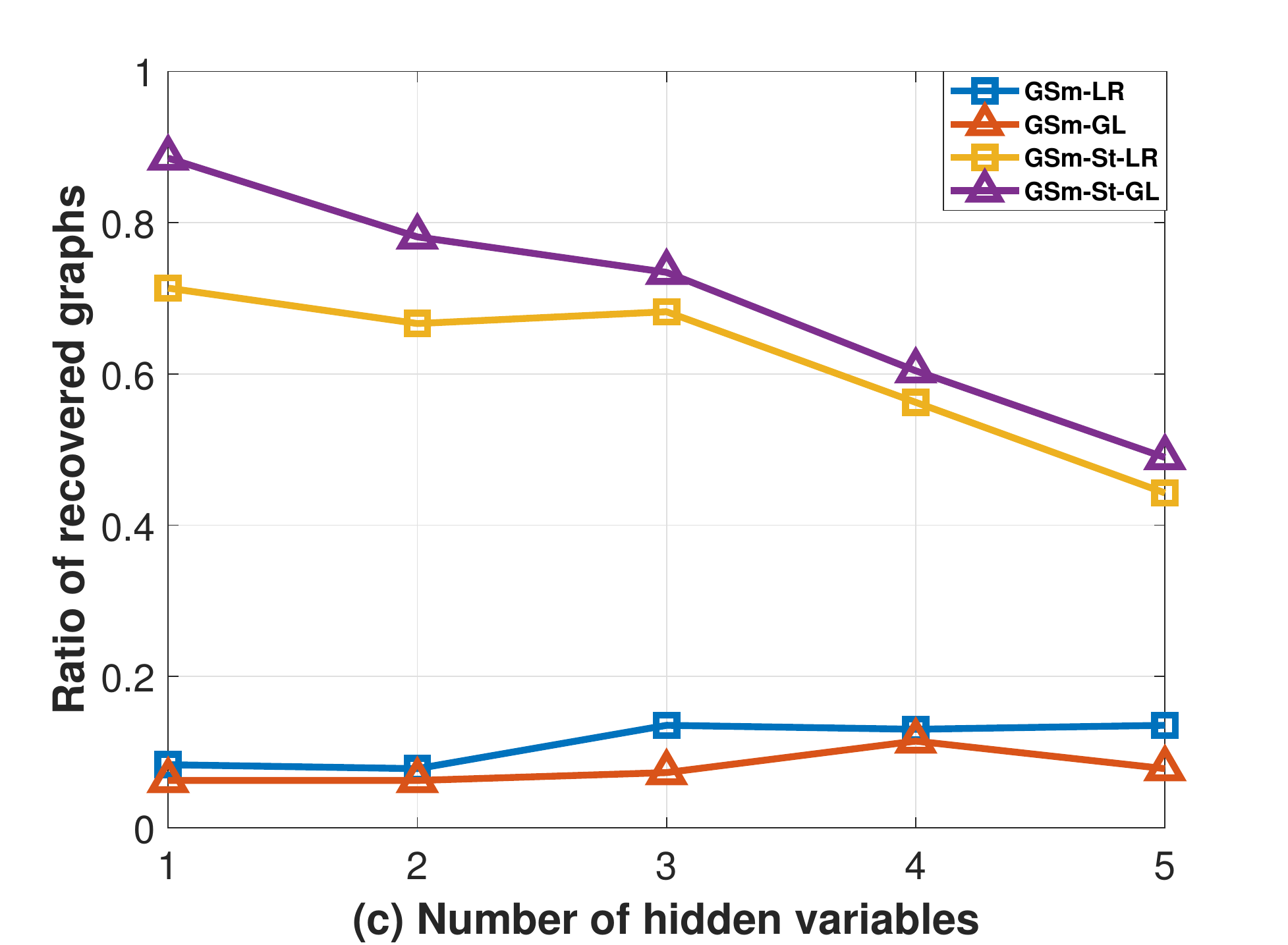}
		%\centering{\small (c)}
	\end{minipage}
	\caption{The ratio of recovered graphs averaged over 200 realizations of random graphs with $N=20$ and stationary observations. The different panels assess the impact of increasing (a) the number of hidden variables $H$ for a scenario with perfectly-known covariance matrices; (b) the number of signal observations $M$ when using the sample covariance matrix; (c) the number of hidden variables $H$ when the inputs are not only stationary but also smooth signals.}
	%%\red{[AGM: Si lo llamamos recory score, hay que hacerlo desde el principio. No lo podemos llamar recovery fraction y, luego de repente, comenzar a usar de forma sistemática una nomenclatura que nunca se ha introducido al lector.]}
	\label{F:exp567}
\end{figure*}

\vspace{0.2cm}
\noindent \textbf{Number of hidden variables.}
This experiment investigates the effect of the hidden nodes on the ability of our algorithms to recover the true graph topology.
To that end, we vary the number of hidden variables $H$.
We consider both the $\bbC_{poly}$ and $\bbC_{MRF}$ models for the observations, assume that the covariance matrices can be perfectly estimated, and select the set of hidden nodes as those with the minimum degree. The results are shown in Figure \ref{F:exp567}, where the $x$-axis represents the number of hidden variables and the $y$-axis the proportion of graphs successfully recovered. The results in Figure~\ref{F:exp567}.a confirm that larger values of $H$ render the inference problem more challenging, leading to a worse ratio of recovered graphs. We also observe that for the $\bbC_{MRF}$ model, LVGL achieves the best performance, especially when $H$ increases. 
This is not surprising since the LVGL is tailored for this specific type of signal generation. On the other hand, LVGL fails to recover any graph when the observed signals follow the more general $\bbC_{poly}$ model. This contrasts with the GSt and GSt-Rw-Fact methods proposed in this paper, which recover the graphs in both settings. 
%This fact makes evident that the stationarity assumption is a more general prior that includes as a specific case the assumption of LVGL, and reflects how making more lenient assumptions entails a more flexible algorithm capable of performing in a wider range of scenarios.
It is also worth noting that the results obtained in Figure~\ref{F:exp567}.a outperform those presented in Figure~\ref{F:exp123}.a. This is due to the fact that graph-stationarity imposes more structure on the observed signals than graph-smoothness, at the expense of needing more observations to accurately estimate the covariance matrices.

\vspace{0.2cm}
\noindent \textbf{Sample covariance matrix.}
The next step is to assess the effect of replacing the true covariance matrix with its sampled estimate $\hbCo=\frac{1}{M}\bbXo\bbXo^\top$.
The number of hidden variables is set to $H=1$, both $\bbC_{MRF}$ and $\bbC_{MRF}$ generative models are tested, the signals are assumed to be Gaussian and zero mean, and all other parameters are set as in the default test-case scenario.
Figure~ \ref{F:exp567}.b illustrates the ratio of recovered graphs as the number of samples $M$ varies. Clearly, the larger the value of $M$ the better the estimate of $\hbCo$.
Analyzing the results in Figure~\ref{F:exp567}.b, we observe that, when using $\bbC_{MRF}$, LVGL obtains the best performance and needs the least number of samples to achieve its best ratio of recovered graphs. 
As noted in the previous experiment, we also observe that LVGL is incapable of recovering graphs when the observations are generated using the $\bbC_{poly}$ model.
On the other hand, GSt and GSt-Rw-Fact achieve a good performance for both covariance models, even though they need a higher number of samples. Finally, GSt-Rw-Fact achieves a performance close to that of LVGL.
This behavior is consistent with the one observed in scenarios where all nodes were observed and latent variables did not exist \cite{segarra2017network}. 
Lastly, upon comparing the results achieved by GSt and GSt-Rw-Fact, the experiments reveal that GSt: i) needs a higher value of $M$ than GSt-Rw-Fact to achieve the same performance, and ii) converges to a worse ratio of recovered graphs.
This is consistent with the results shown in previous experiments and, once again, illustrates the benefits of incorporating additional structure and using more sophisticated regularizers.

\vspace{0.2cm}
\noindent \textbf{Leveraging graph stationarity and smoothness.}
To close the experiments based on synthetic data, we consider here the case where the observed signals are simultaneously smooth and stationary on the unknown graph and evaluate the schemes proposed in Section~\ref{S:smooth_stationary_inf}.
As done in the smooth-based experiments, we create the graph signals as $\bbX = \bbV \bbZ$, with $\bbZ$ sampled from $\ccalN(\bb0,\,\bbLambda^{\dagger})$.
We note that the covariance of $\bbX$ is given by $\bbC = (\bbL^{\dagger})^2$, which is certainly a polynomial of the GSO provided that we set $\bbS=\bbL$. In other words, while the signals generated in Section \ref{Ss:NumExperimentsSmoothSynthetic} were already stationary on the graph, none of the algorithms leveraged that existing structure. Hence, the goal here is to assess the benefits of incorporating that underlying structure into the recovery algorithms.
To that end, we compare the schemes GSm-LR and GSm-GL, which only assume that the signals are smooth on the graph, with GSm-St-LR, which corresponds to \eqref{E:smooth_hidden_st_Lao} with $\gamma_{2,1}=0$, and GSm-St-GL, which corresponds to the \eqref{E:smooth_hidden_st_Lao} with $\gamma_{*}=0$.
Note that GSm-St-LR and GSm-St-GL are, respectively, versions of GSm-LR and GSm-GL that account for the stationarity of the signals.
Figure~\ref{F:exp567}.c shows the ratio of recovered graphs as the number of hidden variables increases for the different algorithms. The advantages of including the stationarity assumption are clear, since, even for $H=3$, the stationary-aware algorithms are able to perfectly recover more than $60\%$ of the generated graphs. 
In contrast, the algorithms that ignore stationarity and account only for smoothness recover correctly less than $20\%$ of the graphs. As expected, including additional information about the observed signals endows the optimization problem with more structure and results in better estimates. If, as in Section \ref{Ss:NumExperimentsSmoothSynthetic}, the recovery performance is measured using the $\Fsco$ associated with individual links, then the differences narrow, with GSm-LR and GSm-GL achieving a (median) $\Fsco$ of around 0.95 and GSm-St-LR and GSm-St-GL a $\Fsco$ that is basically $1$. 

\subsection{Learning graph structure from real datasets}
We close this section by evaluating our algorithms and comparing their recovery performance with existing alternatives in the literature using two real-world datasets.

\vspace{0.2cm}
\noindent \textbf{Learning meteorological graph from temperature data.}
We start by considering the average monthly temperature collected at 88 measuring stations in Switzerland during the period between 1981 and 2010 \cite{temperaturesswiss}.
This leads to a set of signals $\bbX \in \mathbb{R}^{88 \times 12}$, with 12 signals that represent the monthly average temperatures measured at the 88 weather stations.
The goal of the experiment is to use these observations to infer a graph where stations with similar temperature patterns across the year are connected.
While using the geographical graph based on physical distances between the stations can be a more natural (non-data-based) solution to the problem at hand, one must note that Switzerland is a steep terrain. As a result, two nearby stations do not necessarily record similar temperatures across the year, since, for instance, their difference in altitude is large. Motivated by this and, as also done in~\cite{DongLaplacianLearning}, we build the ``ground-truth'' graph upon considering the similarity between stations in terms of their altitude. More specifically, in this experiment, we consider that two stations are connected with a unitary weight if their altitude difference is smaller than 300 meters.
As we want to infer the best-represented graph from the available smooth signals and also take into account the presence of hidden variables, we are going to assume that $\ccalO=\{1,...,20\}$, so that only the 20 first stations are observed, with our goal being inferring the connections between those stations.

We leverage the schemes developed in Section IV (GSm-LR and GSm-GL) and Section VI (GSm-St-LR and GSm-St-GL) to learn the graph associated with the observed nodes from the temperature measurements.
To facilitate comparisons, the evaluation metrics used here are the same as those in~\cite{DongLaplacianLearning}, namely $\Fsco$, $precision$, $recall$, and normalized mutual information (NMI); in addition, the GL-SigRep algorithm from~\cite{DongLaplacianLearning} is used as a baseline. 
The results achieved by the optimal setting of the regularization constants for each of the algorithms are listed in Table \ref{table:1}. 
The main observation is that the explicit consideration of hidden variables when inferring the graph structure leads to better performance. Furthermore, we also observe that GSm-LR outperforms both GL-Sig-Rep and GSm-GL. 
It is also worth noticing that GSm-St-LR and GSm-St-GL obtain the same performance as GSm-LR, revealing that assuming stationarity for this dataset does not seem to further enhance the recovery results.
Although this contrasts with the results from the synthetic experiments, it is not surprising since the number of available samples ($M=12$) is smaller than the number of nodes, which leads to a rank-deficient $\hbCo$ and renders the commutativity constrain inefficient. Indeed, the fact of the covariance being rank-deficient was the reason for not testing the algorithms developed in Section \ref{S:smooth_stationary_inf} in this experiment. 

\begin{table}[]
\caption{Performance achieved by the schemes GL-SigRep (\cite{DongLaplacianLearning}), GSm-LR (Section IV), GSm-GL (Section IV), GSm-St-LR (Section VI) and GSm-St-GL (Section VI) when learning a meteorological graph.}
\centering
\begin{tabular}{lllll}
\hline \hline
%\textbf{Dong}         & 0.8939 & 0.8676    & 0.9219 & 0.6029 \\ \hline
Algorithms                 & $\Fsco$ & Precision & Recall & NMI   \\ \hline \hline 
\textbf{GL-SigRep}        & 0.8800 & 0.9016    & 0.8594 & 0.5746 \\ 
\textbf{GSm-GL}            & 0.9118 & 0.8611    & 0.9688 & 0.6647 \\ 
\textbf{GSm-LR}            & 0.9130 & 0.8514    & 0.9844 & 0.6806 \\ 
\textbf{GSm-St-LR}         & 0.9130 & 0.8514    & 0.9844 & 0.6806 \\ 
\textbf{GSm-St-GL}         & 0.9130 & 0.8514    & 0.9844 & 0.6806 \\ \hline
\end{tabular}

\vspace{-0.3cm}
\label{table:1}
\end{table}

\begin{figure}
%\begin{minipage}[c]{.45\textwidth} %.33 .27
	\includegraphics[width=0.4\textwidth]{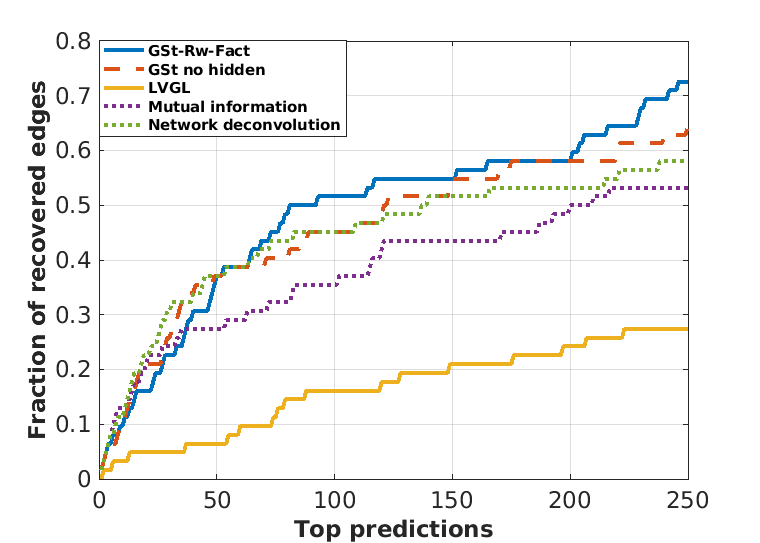}
	\centering
	\vspace{-0.1cm}
	\caption{ Fraction of the real contact edges between amino-acids \cite{Marks2011proteins} recovered for each method as a function of the number of edges considered.}
\vspace{-0.15cm}
%\end{minipage}%}
	\label{F:exp9}
\end{figure}

\vspace{0.2cm}
\noindent \textbf{Learning structural properties of proteins.}
In this case, our goal is to identify the structural properties of proteins from a mutual information graph of the co-variation of amino-acid residues simulating the presence of hidden variables.
We have access to the mutual information matrix of protein BPT1 BOVIN and also to the binary ground-truth contact network built by medical experts, see \cite{FeiziNetworkDeconvolution} and \cite{Marks2011proteins} for details.
The original dimension of both matrices is $53 \times 53$, but in our hidden-variable setup, we consider that we can only observe a submatrix of size $41 \times 41$ and leave the other $12$ nodes as hidden. 
The $y$-axis in Figure \ref{F:exp9} represents the fraction of the real contact edges recovered for several schemes and the $x$-axis represents the number of top-edge predictions. This way, a fraction of recovered edges of 0.6 indicates that if we consider the estimated 100 links with the highest weight, 60 of them match the ground truth links. Five different algorithms are considered: GSt-Rw-Fact (Section \ref{S:smooth_stationary_inf}); GSt no hidden (which approaches the topology-identification problem with stationarity assumptions but ignoring the presence of hidden variables~\cite{marques2016stationaryTSP16}); LVGL; network deconvolution \cite{FeiziNetworkDeconvolution}; and mutual information, with the last two being baselines that have been advocated for this particular dataset. The best performance is achieved by the scheme GSt-Rw-Fact that is accounting for the presence of hidden variables, showcasing the benefits of a more robust formulation. Interestingly, we also observe that even though LVGL accounts for hidden variables, it leads to the worst recovery performance, illustrating the relevance of using topology-inference algorithms that go beyond classical graphical models when dealing with real datasets.  

%%%%%%%%%%%%%%%%%%%%%%%%%%%%%%%%%%%%%%%%%%%%%%%%%%%%%%%%%%%%%%%%%%%%%%%%%%%%%%%%%%%%%%%%%%%%%%%%%%%%%%%%%%%%%%%%%%%%%%%%%%%
%SECTION: CONCLUDING REMARKS
%%%%%%%%%%%%%%%%%%%%%%%%%%%%%%%%%%%%%%%%%%%%%%%%%%%%%%%%%%%%%%%%%%%%%%%%%%%%%%%%%%%%%%%%%%%%%%%%%%%%%%%%%%%%%%%%%%%%%%%%%%%
\section{Conclusions}\label{S:conclusions}
This paper analyzed the problem of inferring the topology of a network from nodal signal observations in the presence of hidden (latent) nodes. To approach this ill-conditioned network-inference task, we considered that the observed signals were (i) smooth on the sought graph; (ii) stationary on the graph; and (iii) a combination of the two previous assumptions.
To render the problem tractable, we further assumed that the number of hidden variables was much smaller than the number of observed nodes and formulated constrained optimization problems that accounted for the topological and signal constraints.
The key to handle the presence of hidden nodes was to consider block-matrix factorization approaches that led to sparse and low-rank constrained optimizations. Since several of the resulting formulations were non-convex, novel judicious convex relaxations were designed.
The performance of the developed algorithms was evaluated in several synthetic and real-world datasets and the results were compared with alternatives from the literature.
%%%%%%%%%%%%%%%%%%%%%%%%%%%%%%%%%%%%%%%%%%%%%%%%%%%%%%%%%%%%%%%%%%%%%%%%%%%%%%%%%%%%%%%%%%%%%%%%%%%%%%%%%%%%%%%%%%%%%%%%%%%

\appendices
\section{Proof of Proposition 1}
Key to our proof are the results from \cite{blockalternating}, which guarantee convergence of BSUM algorithms to a stationary point.

We aim to show that our proposed algorithm satisfies the conditions specified in \cite[Th. 1b]{blockalternating}.
To that end, let $f(\bby)$ represent the objective function in \eqref{E:eqn_factorized}, with $\bby:=[\bby_1^\top,\bby_2^\top,\bby_3^\top]^\top$ and $\bby_1:=\vvec(\bbSo)$, $\bby_2:=\vvec(\bbSoh)$, $\bby_3:=\vvec(\bbCoh)$ denoting the 3 blocks of variables considered in our algorithm.
For each of the $B=3$ block of variables $\bby_b$, we approximate $f(\bby)$ by defining the functions $u_1(\bby_1)$, $u_2(\bby_2)$, and $u_3(\bby_3)$, corresponding to the objective functions in \eqref{E:factorized_step1}, \eqref{E:factorized_step2} and  \eqref{E:factorized_step3}, respectively.
Also, recall that $\ccalY^*$ denotes the set of stationary points of $f(\bby)$ and that $\bby^{(t)}:=[(\bby_{1}^{(t)})^\top,(\bby_{2}^{(t)})^\top,(\bby_{3}^{(t)})^\top]^\top$ is the solution obtained after running $t$ iterations of our algorithm.

With the previous definitions in place, the assumptions required to ensure converge of our algorithm are the following.

\noindent\textit{(\textbf{AS A})} The approximation functions $u_b(\bby_b)$ must be a global upper bound of $f(\bby)$ and the first order behavior of $u_b(\bby_b)$ and $f(\bby)$ must be the same.

\noindent\textit{(\textbf{AS B}) The function $f(\bby)$ must be regular (cf. \cite{blockalternating}) at every point in $\ccalY^*$.} 

\noindent \textit{(\textbf{AS C}) The level set $\ccalY^{(0)} = \{\bby \; | \; f(\bby) \leq f(\bby^{(0)}) \}$ is compact.}

\noindent\textit{(\textbf{AS D}) The problems in 
\eqref{E:factorized_step1}-\eqref{E:factorized_step3} must have a unique solution for any point $\bby^{(t)} \in \ccalY^*$ for at least two of the blocks.}

\noindent We address each of the four assumptions separately, proving that our approach satisfies all of them.

Assumption (\textbf{\textit{AS A}}) requires the surrogate functions $u_b(\bby_b)$ to be global upper bounds of $f(\bby)$.
For the first block ($b=1$), we approximate $f(\bby)$ with the
Taylor series of order 1 of the logarithmic penalty, given by
\begin{align}
    \tilde{u}_1(\bby_1) &=  \sum_{i=1}^{O^2}\log\left(|[\bby_1^{(t)}]_i|+\delta\right) \\
    &+ \sum_{i=1}^{O^2}\frac{\sign([\bby_1^{(t)}]_i)}{|[\bby_1^{(t)}]_i|+\delta}\left([\bby_1]_i-[\bby_1^{(t)}]_i\right) + \rho f_c(\bby_1), \nonumber
\end{align}
where $f_c$ denotes the commutativity penalty in \eqref{E:factorized_step1}.
Since the entries of $\bby_1^{(t)}$ are always either positive or negative [cf. \eqref{E:A_set} and \eqref{E:L_set}], we have that $\sign([\bby_1^{(t)}]_i)[\bby_1]_i=|[\bby_1]_i|$.
After dropping the constant terms, we obtain
\begin{equation}
    u_1(\bby_1) = \sum_{i=1}^{O^2}\frac{|[\bby_1]_i|}{|[\bby_1^{(t)}]_i|+\delta}+ \rho f_c(\bby_1),
\end{equation}
which is the objective function in \eqref{E:factorized_step1}.
%Note that optimizing $\tilde{u}_1$ or $u_1$ over $\bby_1$ equivalent.
Because the $\log$ is a concave differentiable function it follows that its Taylor series of order one constitutes a global upper bound. Therefore, $u_1$ satisfies (\textbf{\textit{AS A}}).
The proof for $u_2$ is equivalent to the proof for $u_1$ so it is omitted for brevity.
Lastly, $u_3(\bby_3)=f(\bby)$ when the blocks $\bby_1$ and $\bby_2$ remain constant, so it also satisfies the requirements, and hence, (\textbf{\textit{AS A}}) is fulfilled.

To proof (\textit{\textbf{AS B}}), according to the definition of regular functions presented in \cite{blockalternating}, it suffices to show that the non-smooth parts of $f(\bby)$ are separable across the different blocks of variables.
To that end, we recall that $\bby_1:=\vvec(\bbSo)$, $\bby_2:=\vvec(\bbSoh)$ and $\bby_3:=\vvec(\bbCoh)$, and decompose $f$ as $f = g_A+g_B+g_C$, with functions $g_A$, $g_B$ and $g_C$ being defined as
\begin{itemize}
    \item $g_A(\bbSo,\bbSoh,\bbCoh) \! = \eta\|\bbSoh\|_F^2 + \eta\|\bbCoh\|_F^2 + \rho\|\hbCo\bbSo\!+\!\bbCoh\bbSoh^\top\!-\!\bbSo\hbCo\!-\!\bbSoh\bbCoh^\top\|_F^2$, where $g_A$ is a smooth function,
    \item $g_B(\bbSo) = \sum_{i,j=1}^{O}\log(|[\bbSo ]_{ij}|+\delta)$, where $g_B$ is a non-smooth function,
    \item $g_C(\bbSoh) = \sum_{i,j=1}^{O,H}\log(|[\bbSoh ]_{ij}|+\delta)$, where $g_C$ is a non-smooth function.
\end{itemize}
Since the non-smooth terms appear in $g_B(\bbSo)$, which only involves variables of the first block $\bby_1=\vvec(\bbSo)$, and $g_C(\bbSoh)$, which only involves variables of the second block $\bby_2=\vvec(\bbSoh)$, it follows that the function $f(\bby)$ is regular for all feasible points.

Next, we show that the level set $\ccalY^{(0)} = \{\bby \; | \; f(\bby) \leq f(\bby^{(0)}) \}$ is compact as required by (\textit{\textbf{AS C}}).
First, note that the entries of $\bbSo$ and $\bbSoh$ are continuous subsets of $\reals$  (e.g., $[\bbSo]_{ij}, [\bbSoh]_{ij} \in \reals_+$ when $\ccalS = \ccalA$), and that $\bbCoh \in \reals^{O \times H}$, so $f(\bby)$ is continuous.
Moreover, since we have that $f(\bby)\leq f(\bby^{(0)})$, this implies that the continuous functions $\log(|[\bbSo]_{ij}|+\delta)$, $\log(|[\bbSoh]_{ij}|+\delta)$, and $\|\bbCoh\|_F^2$ are all bounded, rendering the domain of $f(\bby)$ bounded.
Therefore, it follows that the level set $\ccalY^{(0)}$ is compact.

Finally, since the optimization problems in \eqref{E:factorized_step2} and \eqref{E:factorized_step3} are strictly convex, two of the three problems have unique solutions, satisfying (\textit{\textbf{AS D}}) and concluding the proof.

%%%%%%%%%%%%%%%%%%%%%%%%%%%%%%%%%%%%%%%%%%%%%%%%%%%%%%%%%%%%%%%%%%%%%%%%%%%%%%%%%%%%%%%%%%%%%%%%%%%%%%%%%%%%%%%%%%%%%%%%%%%

\bibliographystyle{IEEEtran.bst}
\bibliography{citations}

\end{document}